\documentclass[a4paper,11pt]{article}
\pdfoutput=1
\usepackage{jcappub}
\usepackage{mathtools}
\usepackage{ulem}
\usepackage{bm}
\usepackage{xcolor}
\usepackage{float}
\usepackage{arydshln}
\usepackage{multirow}
\usepackage{cancel}

\allowdisplaybreaks

\renewcommand{\thesection}{\arabic{section}}
\renewcommand{\theequation}{\thesection.\arabic{equation}}

\def\laq{~\raise 0.4ex\hbox{$<$}\kern -0.8em\lower 0.62ex\hbox{$\sim$}~}
\def\gaq{~\raise 0.4ex\hbox{$>$}\kern -0.7em\lower 0.62ex\hbox{$\sim$}~}

\def\beq{\begin{equation}}
\def\eeq{\end{equation}}
\def\bea{\begin{eqnarray}}
\def\eea{\end{eqnarray}}

\def \pa {\partial}

\def \Hcal {\mathcal{H}}

\newcommand{\dd}{\partial}
\newcommand{\ds}{{\slash\hspace{-5pt}\dd}}
\newcommand{\bds}{\overline{{\slash\hspace{-5pt}\dd}}}

\def \g {\bar{g}}
\def \e {\bar{e}}
\def \u {\bar{u}}
\def \k {\bar{k}}
\def \n {\bar{n}}

\def \B {\mathcal{B}}
\def \C {\mathcal{C}}

\def \S {\mathcal{S}}
\def \V {\mathcal{V}}
\def \T {\mathcal{T}}
\def \Vc {\checkmark}

\title{Gauge invariance on the light-cone: curvature perturbations and radiative degrees of freedom}

\author[a]{G. Fanizza,}
\author[b]{G. Marozzi}
\author[b]{and M. Medeiros}

\affiliation[a]{Instituto de Astrof\'isica e Ci\^encias do Espa\c{c}o, Faculdade de Ci\^encias da Universidade de Lisboa, Edificio C8, Campo Grande, P-1749-016, Lisbon, Portugal}
\affiliation[b]{Dipartimento di Fisica, Universit\`a di Pisa, Largo B. Pontecorvo 3, 56127 Pisa, 
Italy,\\
and Istituto Nazionale di Fisica Nucleare, Sezione di Pisa, Italy}

\emailAdd{gfanizza@fc.ul.pt}
\emailAdd{giovanni.marozzi@unipi.it}
\emailAdd{matheus.rodriguesmedeirossilva@phd.unipi.it}

\abstract{We derive the expressions on the observed light-cone for some relevant cosmological gauge invariant variables, such as the Mukhanov-Sasaki variable and $E$- and $B$- modes of the tensor perturbations. Since the structure of the light-cone does not reflect in a direct way the FLRW symmetries, we develop a formalism which is coordinate independent and classifies the perturbations according to their helicities. Even though we work with linear perturbations, our formalism can be readily extended to non-linear theory and put the basis to study the evolution of cosmological perturbations, since the early- until the the late-time Universe, directly along the observed light-cone.}

\keywords{cosmological perturbation theory,
geodesic light-cone gauge, gravitational waves
 
\vskip13pt plus8pt minus11pt

\noindent{\bfseries\large\sffamily{Preprints:}}}

\begin{document}

\maketitle

\section{Introduction}
The Geodesic Light-Cone (GLC) gauge \cite{Gasperini:2011us} has shown several advantages in many aspects of the late-time theoretical cosmology. Few examples are the exact computation of light-like cosmological observables, such as the redshift \cite{Gasperini:2011us,Fanizza:2018qux}, the luminosity distance \cite{BenDayan:2012pp,BenDayan:2012ct,BenDayan:2012wi,Fanizza:2013doa, Marozzi:2014kua}, the Jacobi map \cite{Fanizza:2013doa,Fanizza:2022wob}, the galaxy number count \cite{DiDio:2014lka,DiDio:2015bua} and the evaluation of well-posed averages of these observables \cite{Gasperini:2011us,Fanizza:2019pfp}. Furthermore, the GLC gauge has been successfully applied also to the study of non-linear corrections to CMB spectra \cite{Fanizza:2015swa,Marozzi:2016uob,Marozzi:2016qxl} and to investigate the behavior of ultra-relativistic particles in the inhomogeneous Universe \cite{Fanizza:2015gdn,Fleury:2016mul}.

These advantages have their root in the fact that the GLC gauge is fixed in such a way that the light-like geodesics are exactly solved. Notwithstanding this remarkable feature, a detailed study of the cosmological dynamics in the GLC coordinates \cite{Mitsou:2020czr} has shown how this is quite involved already at linear order, at least for an analytic treatment. 

Because of these facts, a sort of dualism seems to emerge between the standard perturbative description of cosmological inhomogeneities and the fully non-linear light-cone one. The former admits manageable (and explicitly solvable) equation of motions but rather complicated expressions for light-like cosmological observables, whereas the latter has simple non-linear forms for the cosmological observables (and their averages) but a quite cumbersome evolution of the dynamics.

A possible way out of this dichotomy may be provided by numerical attempts. As a relevant example, in a recent work \cite{Tian:2021qgg}, the linearized equation of motion on the light-cone for the gravitational potential has been solved. This open an unprecedented opportunity to study the evolution of perturbations along the observed past light-cone. However, for this program to be successful, a systematic formulation of perturbation theory along the light-cone is needed, in order to properly identify the relevant (combinations of the) degrees of freedom for the quantities of cosmological interest.

In \cite{Fanizza:2020xtv}, we have built a cosmological perturbation theory directly on the light-cone and provided a general road map for the computation of gauge invariant observables. This has been applied to the study of  the distance-redshift relation. Here we aim to extend our approach to describe other gauge invariant degrees of freedom of cosmological interest, such as the curvature perturbation and the tensor ones. As we are going to show, a key ingredient for the evaluation of the latter will be a decomposition of the perturbations based on their behavior as Scalar/PseudoScalar (SPS) under rotation on the sphere rather than the usual Scalar-Vector-Tensor (SVT) decomposition. Thanks to this, we will be able to provide the explicit expression for the $E$- and $B$- mode of the tensor perturbations, which correspond to the physical radiative degrees of freedom of the theory.

Finally, we acknowledge that in \cite{Frob:2021ore} the study of the gauge invariant curvature perturbation in the GLC gauge has also been applied to obtain expressions for the expansion rate in the early universe in order to investigate the backreaction effects along the past light-cone. In this regard, what this work adds to this computation is the explicit gauge invariant expression for this quantity, whereas in \cite{Frob:2021ore} the linearized GLC has been adopted. As a sanity check, we prove that our generalized result agrees with \cite{Frob:2021ore}.

The manuscript is organized as follows. In Sect.~\ref{SPT}, we first recall the standard perturbation theory and set the notation. Afterwards, we report some of the main results of \cite{Fanizza:2020xtv} and provide the connection between the standard SVT perturbations and the SPS ones. In Sect.~\ref{sec:gen}, we introduce a manner to decompose the perturbations based on a set of projections which takes into account the helicity of the perturbations, is coordinate independent and extendable to higher perturbative orders. Also in this section, we apply this formalism to both standard perturbation theory and the light-cone one. Hence, in Sect.~\ref{sec:EB} we extract the $E$- and $B$- modes for the gauge invariant tensor perturbations as function of the light-cone SPS perturbations. Finally, the main conclusions are discussed in Sect.~\ref{sec:SC}. Furthermore, in Appendix~\ref{app:GT} we explicitly check that the relations between SVT and SPS satisfy the expected gauge transformations. Appendix~\ref{app:FL} is devoted to show the agreement with the results and the formalism of \cite{Frob:2021ore}, whereas Appendix~\ref{app:swsh} reports some technical aspects concerning the spin-raising and spin-lowering operators needed along the paper. Appendix~\ref{app:EB} contains the technical details for the evaluation of the $E$- and $B$- modes of the gauge invariant tensor perturbations.

\section{A conventional approach}
\label{SPT}
In this section, we will make contact between the standard perturbation theory and the light-cone one. We will briefly review the standard approach to linear perturbations and the main gauge invariant variables of cosmological interest. Then, we will introduce a suitable set of perturbations for the light-cone structure. Finally, after having developed the appropriate dictionary between the two sets, we will derive gauge invariant quantities such as tensor perturbations and Mukhanov-Sasaki variable in terms of the light-cone perturbations. This work is a natural continuation of the formalism already developed in \cite{Fanizza:2020xtv,Mitsou:2020czr}.

\subsection{Standard Perturbation Theory}
Here, we briefly report the well-known results of linear perturbations of Friedmann-Lema\^itre-Robertson-Walker (FLRW) metric in order to set the notation that will be followed in the rest of the paper. By defining with $g_{\mu\nu}$ the inhomogeneous metric tensor and with $\delta g_{\mu\nu}$ the set of perturbations on top of the FLRW metric $\bar{g}_{\mu\nu}$, we have\footnote{Greek indices run from $0$ to $4$, Latin ones $i,j,k,\dots$ range from $1$ to $3$ and Latin ones $a,b,c,\dots$ label angular coordinates.}
\begin{equation}
g_{\mu\nu}\equiv\bar{g}_{\mu\nu}+\delta g_{\mu\nu}=a^{2}\left[\begin{pmatrix}-1 & 0\\
0 & \bar{\gamma}_{ij}
\end{pmatrix}+\begin{pmatrix}-2\phi & -\mathcal{B}_{i}\\
-\mathcal{B}_{i} & \mathcal{C}_{ij}
\end{pmatrix}\right]\,.\label{eq:pertFLRW}
\end{equation}
In spherical coordinates, the background spatial metric is $\bar{\gamma}_{ij}=\text{diag}\left(1,\bar{\gamma}_{ab}\right)$, the metric on the 2-sphere is $\bar{\gamma}_{ab}=r^2\text{diag}\left(1,\sin^{2}\theta\right)$ and $r$ is the space-like radial coordinate. Perturbations are typically decomposed according to the irreducible representations of the background symmetry groups $SO\left(3\right)$ as
\begin{align}
\mathcal{B}_{i} & =\partial_{i}B+B_{i}\,,
\nonumber\\
\mathcal{C}_{ij} & =-2\,\psi\,\gamma_{ij}+2\,D_{ij}E+2\,\nabla_{(i}F_{j)}+2\,h_{ij}\,,\label{eq:BC}
\end{align}
where the covariant derivative $\nabla_i$ is defined as
$\nabla_{i}S^{j}=\partial_{i}S^{j}+\Gamma_{il}^{j}S^{l}$ with $\Gamma_{il}^{j}$ the Christoffel connection w.r.t. $\bar{\gamma}_{ij}$,  $D_{ij}\equiv\nabla_{(i}\nabla_{j)}-\frac{1}{3}\bar{\gamma}_{ij}\Delta_{3}$ and $\Delta_{3}\equiv\bar{\gamma}^{ij}\nabla_{i}\nabla_{j}$. Eq.~\eqref{eq:BC} is known as Scalar-Vector-Tensor (SVT) decomposition, since $B,\psi,E$ and $\phi$ are scalars, $B_{i}$ and $F_{i}$ are divergence-less vectors ($\nabla^{i}B_{i}=\nabla^{i}F_{i}=0$),
and $h_{ij}$ is a trace-less and divergence-less
tensor ($\bar{\gamma}^{ij}h_{ij}=0$ and $\nabla^{i}h_{ij}=0$) when the metric is transformed according to the above-mentioned symmetry group.

Eq.~\eqref{eq:BC} still admits a gauge freedom in the definition of the perturbations $\delta g_{\mu\nu}$.
This is fixed by exploring the transformation property under diffeomorphism: given a coordinate
transformation $y^{\mu}\rightarrow\widetilde{y^{\mu}}=y^{\mu}+\epsilon^{\mu}$, perturbations $\delta g_{\mu\nu}$ change as
\begin{equation}
\widetilde{\delta g_{\mu\nu}}(y^\mu)=\delta g_{\mu\nu}(y^\mu)
-\mathcal{L}_{\mathbf{\epsilon}}g_{\mu\nu}(y^\mu)\,,
\label{eq:gauge-tranform-metric}
\end{equation}
where $\mathcal{L}_{\mathbf{\epsilon}}(g_{\mu\nu})\equiv\hat{\nabla}_{\mu}\epsilon_{\nu}+\hat{\nabla}_{\nu}\epsilon_{\mu}$ is the Lie derivative of $g_{\mu\nu}$ w.r.t. the field $\epsilon^\mu$ and $\hat{\nabla}_{\mu}$ is the covariant derivative
 w.r.t. the 4-dimensional background metric $\bar{g}_{\mu\nu}$.
The SVT decomposition can be applied to the space-like part of $\epsilon^\mu$ as well. This simply returns $\epsilon^i=e^i+\partial^i\epsilon$. We are then left with two scalars gauge fields $\epsilon^\eta$ and $\epsilon$ and a divergence-less vector gauge field $e^i$. The straightforward combination of Eqs.~\eqref{eq:BC} and \eqref{eq:gauge-tranform-metric} leads to the well-known gauge transformations for scalars
\begin{align}
\widetilde{\phi} =&\,\phi-\partial_{\eta}\epsilon^{\eta}+\mathcal{H}\epsilon^{\eta}\,,
&\widetilde{E}  =&\,E-\epsilon\,,
\nonumber\\
\widetilde{\psi} =&\,\psi+\mathcal{H}\epsilon^{\eta}+\frac{1}{3}\Delta_{3}\epsilon\,,
&\widetilde{B}  =&\,B-\epsilon^{\eta}+\partial_{\eta}\epsilon\,,
\label{eq:gaugetransf}
\end{align}
and vectors
\begin{equation}
\widetilde{B}_{i}=B_{i}+\partial_{\eta}\left(\frac{e_{i}}{a^{2}}\right)\,,
\qquad\qquad
\widetilde{F}_{i} =F_{i}-\frac{e_{i}}{a^{2}}\,,
\label{eq:gaugetransfvector}
\end{equation}
whereas tensor perturbations are gauge invariant $\widetilde{h}_{ij}=h_{ij}$, since there is not any linear tensor gauge mode in $\epsilon^\mu$.

Gauge dependence affects also the matter sector. Indeed, given a scalar $f$, its gauge transformation is given by
 \begin{equation}
     \widetilde{f}(y^\mu)=f(y^\mu)-\mathcal{L}_{\epsilon}f(y^\mu)\,.
     \label{eq:scalartr}
 \end{equation}
By splitting the general scalar field $f$ in a time-dependent background $\bar{f}$ and its linear fluctuations $\delta f$, we will then obtain the following gauge transformation
\begin{equation}
     \widetilde{\delta f}=\delta f -\epsilon^{\eta}\partial_{\eta}\bar{f}\,.
     \label{eq:gt-scalarf}
 \end{equation}
Of particular interest is the case of a single field slow-roll inflationary model, where we have a perfect fluid, with energy-density $\rho$ and pressure $p=w\rho$, and a scalar field $\varphi$. In this case we have
 \begin{align}
 \widetilde{\delta\rho}=&\delta\rho-\epsilon^{\eta}\partial_{\eta}\bar{\rho}\,,\nonumber\\
         \widetilde{\delta\varphi}=&\delta\varphi-\epsilon^{\eta}\partial_{\eta}\bar{\varphi}\,,
         \label{eq:fluidstr}
 \end{align}
where $\delta\rho$ and $\delta\varphi$ are the linear fluctuations of $\bar\rho$ and $\bar\varphi$ respectively.

We can then get rid of the gauge dependence by building linear combinations of perturbations. Concerning the scalar sector, few relevant examples are given by the following gauge invariant variables. We can work exclusively with the metric perturbations, where particularly important variables are the so-called Bardeen potentials
\begin{align}
\Psi = &\,\psi+\mathcal{H}\left(B+\partial_{\eta}E\right)+\frac{1}{3}\Delta_{3}E\,,
\nonumber\\
\Phi = &\,\phi-\partial_{\eta}\left(B+\partial_{\eta}E\right)-\mathcal{H}\left(B+\partial_{\eta}E\right)\,.
\label{eq:Bardeen}
\end{align}
Other realizations of gauge invariant quantities relate matter to geometry and are provided by
\begin{align}
\zeta & =\psi+\frac{1}{3}\Delta_{3}E-\frac{1}{3}\frac{\delta\rho}{\rho+p}\,,
\nonumber\\
\mathcal{R} & =\psi+\frac{1}{3}\Delta_{3}E+\left(\frac{\mathcal{H}}{\partial_{\eta}\bar\varphi}\right)\delta\varphi
\,,
\label{eq:GI-quantities}
\end{align}
which are both gauge invariant expressions related to the curvature perturbations. Indeed, we recognize the combination $\psi+\frac{1}{3}\Delta_3E$ as the {\it gauge dependent curvature perturbation} which, as well-known, transforms with a temporal gauge mode.
The different gauge fixings for this gauge mode in the curvature perturbation then play an important role in the current cosmological paradigm. Indeed $\zeta$ represents the gauge invariant quantity whose value equals the spatial curvature perturbation as evaluated on constant uniform energy density hypersurfaces. In the same manner, $\mathcal{R}$ represents the spatial curvature perturbation evaluated on hypersurfaces where the scalar field $\varphi$ is uniform. This last one defines the physical curvature perturbations in the primordial Universe which has to be connected with late-time inhomogeneities, since the inflaton is the only clock available for single field models. Such variable is then directly connected with the so-called Mukhanov-Sasaki variable \cite{Mukhanov:1985rz,Mukhanov:1988jd,Sasaki:1986hm} given by 
\begin{equation}
    Q=\frac{\partial_{\eta}\bar{\varphi}}{\mathcal{H}}
    \mathcal{R}\,.
\label{eq:GI-quantities-Q}
\end{equation}

For a later use, let us see the way to extract the
SVT parameters from Eqs.~\eqref{eq:BC}. Starting
from Eqs.~\eqref{eq:BC}, we can first extract the scalar perturbation $B$ by applying $\nabla^{i}$
to $\mathcal{B}_i$. We then have that
\begin{equation}
\Delta_3 B =\,\nabla^{i}\mathcal{B}_{i}
\qquad\text{and}\qquad
B_{i} =\,\mathcal{B}_{i}-\partial_{i}\Delta^{-1}_3\left(\nabla^{j}\mathcal{B}_{j}\right)\,.
\label{eq:B}
\end{equation}
In a similar manner, we can express the SVT perturbations in $\mathcal{C}_{ij}$ from \eqref{eq:BC}. By projecting $\mathcal{C}_{ij}$ onto $\bar{\gamma}^{ij}$ and $D^{ij}$, we get the scalars
\begin{equation}
\psi=-\frac{1}{6}\bar{\gamma}^{ij}\mathcal{C}_{ij}\qquad\text{and}\qquad
\left(\Delta_{3}\right)^{2}E=\frac{3}{4}D^{ij}\mathcal{C}_{ij}\,.\label{eq:laplE-1}
\end{equation}
Hence, by using Eqs.~\eqref{eq:laplE-1}, we can extract $F_{i}$ from $\nabla^j\mathcal{C}_{ij}$, and obtain
\begin{equation}
\Delta_3F_{i}=\nabla^{j}\mathcal{C}_{ij}+6\nabla_{i}\psi-4\nabla_{i}\left(\psi+\frac{1}{3}\Delta_{3}E\right)\,,\label{eq:Fi-suitable}
\end{equation}
where we split the contribution of $\psi$ in order to reconstruct the gauge dependent curvature in the last term.
Finally, thanks to Eqs.~\eqref{eq:laplE-1} and \eqref{eq:Fi-suitable},
we have that the gauge invariant tensor perturbations are
\begin{equation}
h_{ij}=\frac{1}{2}\mathcal{C}_{ij}-\frac{3\nabla_{i}\nabla_{j}}{\Delta_{3}}\psi+\left[\gamma_{ij}+\frac{\nabla_{i}\nabla_{j}}{\Delta_{3}}\right]\left(\psi+\frac{1}{3}\Delta_{3}E\right)-\frac{1}{\Delta_{3}}\left[\nabla^{l}\nabla_{(i}\mathcal{C}_{j)l}\right]\,.\label{eq:h_ij}
\end{equation}
We remark that $h_{ij}$ in Eq.~\eqref{eq:h_ij} is defined up to a free-function $f$ such that $\Delta_3 f=0$. This follows from the fact that we needed to invert the Laplacian in Eqs.~\eqref{eq:laplE-1} and \eqref{eq:Fi-suitable}.

The SVT decomposition here provided simply makes use of the irreducible representations in terms of the $SO(3)$ group. This is a symmetry of the background, once a foliation of the manifold in terms of constant-time hypersurfaces is adopted, where also the operators $\bar{\gamma}_{ij}$, $D_{ij}$, $\nabla_{i}$ are defined. However, no assumptions about a perturbative expansion for $\delta g_{\mu\nu}$ has been required in order to derive Eqs.~\eqref{eq:B}-\eqref{eq:h_ij}. Therefore, all these equations can be interpreted as a non-linear decomposition in the following sense:
the standard SVT decomposition used in Eqs.~\eqref{eq:B}-\eqref{eq:h_ij} is based on the background
symmetries, allowing then to classify Scalar, Vector and Tensor perturbations 
according to a finite 3-D rotation of the spatial coordinates which is allowed by the
$SO(3)$ symmetry of the FLRW background. For the covariant form of the
metric tensor, then the decomposition in Eq.~\eqref{eq:BC} is valid to any perturbative order and this is the
only thing that we need to evaluate Eqs.~\eqref{eq:B}-\eqref{eq:h_ij} . In this sense, these relations
can be readily thought as non-linear expressions. Clearly, this is no longer the case
when we invert the metric to the contro-variant form where, indeed, the non-linear
expressions become quite complicated.

\subsection{Light-Cone Perturbations}
\label{GLC}

Here we will discuss a set of perturbations better adapted to the structure of the light-cone. We will follow the formalism already developed in \cite{Fanizza:2020xtv}.

To this end, we start from the Geodesic Light-Cone (GLC) gauge. This is a fully non-linear metric expressed in terms of the light-cone coordinates $x^{\mu} = \left(\tau, \, w, \, \theta^{a} \right) $, where $\tau$ is the proper time of a geodetic observer with four-velocity
$u_ {\mu} = -\partial_ {\mu}\tau = -\delta_ {\mu}^{\tau} $, $w$ is a null coordinate which describes
the observer's past light-cone and $\theta^{a}$ label the incoming directions of light-like signals in the observer's frame. This coordinates system is built in a manner that both time-like geodesics and light-like geodesics are exactly solved respectively by
$u_ {\mu} = -\delta_ {\mu}^{\tau} $ and $k_ {\mu} = \partial_ {\mu} w = \delta_ {\mu}^{w} $. This is possible thanks to the fixing of the gauge freedom allowed by general relativity, rendering then the fully inhomogeneous metric tensor as \cite{Gasperini:2011us}
\begin{align}
g^{GLC}_{\mu\nu}=\begin{pmatrix}0 & -\Upsilon & \vec{0}\\
-\Upsilon & \Upsilon^{2}+U^{2} & -U_{a}\\
\vec{0}^\mathbf{T} & -U_{a}^{\mathbf{T}} & \gamma_{ab}
\end{pmatrix}\,.
\label{eq:GLCmetric}
\end{align}
As shown in \cite{Fanizza:2020xtv}, the linearization of \eqref{eq:GLCmetric} corresponds to the following gauge fixing in standard perturbation theory in Eq.~\eqref{eq:pertFLRW}
\begin{equation}
\phi=0\qquad,\qquad\mathcal{B}_{r}=-\frac{C_{rr}}{2}\qquad,\qquad\mathcal{B}_{a}=-\mathcal{C}_{ar}\,,\label{eq:StandardGLCcond}
\end{equation}
and this gauge fixing has been dubbed as \textit{Observational Synchronous Gauge} (OSG).

Now, in order to apply the SPS decomposition and its classification for the gauge modes, we start from the homogeneous FLRW-like background metric \eqref{eq:GLCmetric}. This is given by the condition $\Upsilon=a$, where
$a(\tau)$ is the scale factor and $U_{a}=0$. The latter just follows from the isotropy of the background, where there is no misalignment of the light-geodesics between two light-cones in the FLRW geometry
(differently from anisotropic geometries, as shown in \cite{Fleury:2016htl}).
Additionally, we take
\begin{equation}
d\tau=ad\eta
\qquad\text{and}\qquad
dw=dr+d\eta
\label{eq:2.18}
\end{equation}
to link $\tau$ and $w$ to $\eta$ and $r$, whereas the angles remain unaltered, i.e. $d\theta^a_{FLRW}=d\theta^a_{GLC}$\footnote{Thanks to this equality, in the following we will omit the label FLRW or GLC when referring to the angular part of the metric.}. In this way, the desired framework for the light-cone perturbations is~\cite{Fanizza:2020xtv}
\begin{equation}
f_{\mu\nu}=\bar{f}_{\mu\nu}
+\delta f_{\mu\nu}=
\begin{pmatrix}0 & -a & \vec{0}\\
-a & a^{2} & \vec{0}\\
\vec{0}^\mathbf{T} & \vec{0}^\mathbf{T} & a^{2}\bar{\gamma}_{ab}
\end{pmatrix}+a^{2}\begin{pmatrix}L & M & V_{a}\\
M & N & U_{a}\\
V^\mathbf{T}_{a} & U^\mathbf{T}_{a} & \delta\gamma_{ab}
\end{pmatrix}\,.\label{eq:generalGLCpert}
\end{equation}
Given that $U_a$ in Eq.~\eqref{eq:generalGLCpert} is a purely linear quantity, since now on, it will be meant as a perturbation, unless differently specified.

The angular perturbations in $\delta f_{\mu\nu}$ may be further decomposed in their irreducible form as 
\begin{align}
V_{a}=&\,r^{2}\left(D_{a}v+\widetilde{D}_{a}\hat{v}\right)\,,\nonumber\\
U_{a}=&\,r^{2}\left(D_{a}u+\widetilde{D}_{a}\hat{u}\right)\,,\nonumber\\
\delta\gamma_{ab}=&\,2r^{2}\left(\bar{q}_{ab}\nu+D_{ab}\mu+\widetilde{D}_{ab}\hat{\mu}\right)\,,
\label{eq:sphere-decomp}
\end{align}
where $D_{a}$ is the covariant derivative w.r.t. the metric on the sphere $\bar{q}_{ab}=\text{diag}\left(1,\sin^{2}\theta^1\right)$, $\widetilde{D}_{a}=\epsilon_{a}^{\,b}D_{b}$,
where $\epsilon_{ab}$ is the Levi-Civita symbol, $D_{ab}\equiv D_{(a}D_{b)}-\frac{1}{2}\bar{q}_{ab}D^2$ and $\widetilde{D}_{ab}=D_{(a}\widetilde{D}_{b)}$.

Perturbations are classified as scalars $v,\,u,\,\mu$ and $\nu$, and pseudo-scalar $\hat{v},\,\hat{u}$ and $\hat{\mu}$ under a rotation on the 2-D sphere where $\tau$ and $w$ are constants. We underline that this perturbative scheme has been built on a background where the light-cone structure is encoded. Indeed, as proven in \cite{Mitsou:2020czr}, the exact GLC gauge is equivalent to a 1+1+2 ADM-like foliation of the Lorentzian manifold, where the 1+1 subspace contains a time-like and a light-like hypersurface. On the other hand, the remaining 2-D space-like hypersurface can be interpreted as a set of spheres at constant time on the same light-cone. In our case, this structure precisely reflects into the background metric $\bar{f}_{\mu\nu}$.

Here we anticipate that the trace-less perturbations of the 2-D sphere $\mu$ and $\hat{\mu}$ describe purely spin-2 gauge invariant perturbations in the helicity basis. In fact, as we will also show along this work, and previously claimed in \cite{Mitsou:2019nhj, Mitsou:2020czr, Fanizza:2020xtv}, the scalar/pseudo-scalar decomposition presented in \eqref{eq:sphere-decomp} is most suitable to describe $E$ and $B$ modes polarization of the perturbations in terms of the helicity decomposition (see also \cite{Bernardeau:2009bm,Schmidt:2012ne,Hu:2000ee} and the references therein, and \cite{Fanizza:2022wob} for applications to the GLC gauge).

\subsection{SVT-GLC relation}
\label{SVT-GLC}

In this subsection, we will make contact between the standard SVT decomposition and the linear light-cone perturbations just presented.
To this end, we first report the relations between $\delta f_{\mu\nu}$ and $\delta g_{\mu\nu}$ already found in Eqs.~(3.7) of \cite{Fanizza:2020xtv}. Indeed, thanks to the coordinate transformation  \eqref{eq:2.18}, we have
\begin{align}
\phi=&-\frac{1}{2}\left(a^{2}L+N+2aM\right)\,,
&\mathcal{B}_{r} & =-\left(N+aM\right)\,,
&\mathcal{B}_{a} & =-\left(U_{a}+aV_{a}\right)\,,\nonumber\\
\mathcal{C}_{rr} & =N\,,
&\mathcal{C}_{ra} & =U_{a}\,,
&\mathcal{C}_{ab} & =\delta\gamma_{ab}\,.
\label{eq:Crr}
\end{align}
Now we can use the results from Sect.~\ref{GLC} and relate the SVT decomposition to the GLC perturbations. 

This can be done by extracting the SVT parameters directly from operators appearing within the GLC d.o.f., once the decomposition in Eqs.~\eqref{eq:BC} is adopted. Hence, by inserting Eqs.~\eqref{eq:Crr} into Eqs.~\eqref{eq:B}-\eqref{eq:h_ij}, we have that the scalars are
\begin{align}
\psi=&-\frac{1}{6}\left(N+4\nu\right)\,,\nonumber\\
\Delta_3B
=&-\left(\partial_{w}+\frac{2}{r}\right)\left(N+aM\right)-D^2\left(u+av\right)\,,\nonumber\\
\left(\Delta_{3}\right)^{2}E
  =&\,\frac{1}{2}\left(\partial_{w}^{2}+\frac{5}{r}\partial_{w}+\frac{3}{r^{2}}\right)\left(N-2\nu\right)-\frac{1}{4\,r^2}D^{2}\left(N+4\nu\right)\nonumber\\
 & +\frac{1}{6\,r^{2}}\left(\partial_{w}+\frac{1}{r}\right)D^2u+\frac{3}{2\,r^2}D^2\nu
 +\frac{3}{4\,r^2}\left(D^2\right)^2\mu +\frac{3}{2r^{2}}D^{2}\mu\,,
\label{eq:scal}
\end{align}
where we made use of the following useful relations
\begin{align}
\nabla^{i}\mathcal{C}_{ir}
=&\left(\partial_{w}+\frac{2}{r}\right)N+D^{2}u-\frac{4}{r}\nu\,,\nonumber\\
\nabla^{i}\mathcal{C}_{ia}
=&\left(\partial_{w}+\frac{2}{r}\right)\left[r^{2}\left(D_{a}u+\tilde{D}_{a}\hat{u}\right)\right]+2D_{a}\nu\,
+D_{a}\left(D^{2}\mu+2\mu\right)+\tilde{D}_{a}\left(D^{2}\hat{\mu}+2\hat{\mu}\right)\,,\nonumber\\
\nabla^{i}\nabla^{j}\mathcal{C}_{ij}
 =&\,\partial_{w}^{2}N+\frac{4}{r}\left(\partial_{w}+\frac{1}{2r}\right)N+\frac{2}{r^{2}}\left(\partial_{w}+\frac{1}{r}\right)\left(r^{2}D^{2}u\right)\,\nonumber\\
 &-\frac{4}{r}\left(\partial_{w}+\frac{1}{r}\right)\nu+\frac{2}{r^{2}}D^{2}\nu\
 +\frac{1}{r^{2}}\left(D^{2}+2\right)D^{2}\mu\,.
 \label{eq:ninjcij}
\end{align}
In the same way, we find that the vector field $B_i$ is
\begin{align}
\Delta_3B_{r}
=& -\Delta_3\left(N+aM\right)
+\partial_{w}\left[\left(\partial_{w}+\frac{2}{r}\right)\left(N+aM\right)+r^{2}D^{2}\left(u+av\right)\right]\,,\nonumber\\
\Delta_3B_{a}
=&\,\Delta_3\left[r^{2}D_{a}\left(u+av\right)+r^{2}\tilde{D}_{a}\left(\hat{u}+a\hat{v}\right)\right]\,\nonumber\\
&+\partial_{a}\left[\left(\partial_{w}+\frac{2}{r}\right)\left(N+aM\right)+r^{2}D^{2}\left(u+av\right)\right]\,.\label{eq:Ba-GLC}
\end{align}

Finally, by inserting Eqs.~\eqref{eq:scal} and \eqref{eq:ninjcij}
into Eq.~\eqref{eq:Fi-suitable}, we also express the components of the vector field $F_{i}$ in terms of GLC perturbations as
\begin{align}
\Delta_3F_{r}
=& -\partial_{w}\left(N+4\nu\right)
+\left(\partial_{w}+\frac{2}{r}\right)N+D^{2}u-\frac{4}{r}\nu\nonumber\\
&- \frac{4\partial_{w}}{\Delta_{3}}\left[ \frac{1}{2}\left(\frac{1}{r}\partial_{w}+\frac{1}{r^{2}}-\frac{1}{2r^{2}}D^{2}\right)N+\frac{1}{4r^{2}}\left(D^{2}\right)^{2}\mu+\frac{1}{2r^{2}}D^{2}\mu\right.\,\,\nonumber\\
&+ \left.\frac{1}{2}\left(\partial_{w}+\frac{3}{r}\right)D^{2}u-\left(\partial_{w}^{2}+\frac{3}{r}\partial_{w}+\frac{1}{r^{2}}+\frac{1}{2r^{2}}D^{2}\right)\nu\right]\,,\nonumber\\
\Delta_3F_{a}
 =& -\partial_{a}\left(N+4\nu\right)
+\left(\partial_{w}+\frac{2}{r}\right)\left[r^{2}\left(D_{a}u+\tilde{D}_{a}\hat{u}\right)\right]
+2D_{a}\nu\,\nonumber\\
&+D_{a}\left(D^{2}\mu+2\mu\right)+\tilde{D}_{a}\left(D^{2}\hat{\mu}+2\hat{\mu}\right)\,\nonumber\\
 &-\frac{4\partial_{a}}{\Delta_{3}}\left[ \frac{1}{2}\left(\frac{1}{r}\partial_{w}+\frac{1}{r^{2}}-\frac{1}{2r^{2}}D^{2}\right)N+\frac{1}{4r^{2}}\left(D^{2}\right)^{2}\mu+\frac{1}{2r^{2}}D^{2}\mu\right.\,,\,\nonumber\\
&+ \left.\frac{1}{2}\left(\partial_{w}+\frac{3}{r}\right)D^{2}u-\left(\partial_{w}^{2}+\frac{3}{r}\partial_{w}+\frac{1}{r^{2}}+\frac{1}{2r^{2}}D^{2}\right)\nu\right]\,.
\label{eq:Fa}
\end{align}

What emerges from Eqs.~\eqref{eq:scal}-\eqref{eq:Fa} is that the standard scalar and vector perturbations look quite involved once expressed in terms of the light-cone perturbations, even though their gauge transformation is consistent with the ones expected in the SVT approach (see Appendix \ref{app:GT} for the explicit proves). Before discussing how this procedure is applied to the tensor perturbations and to other gauge invariant variables of cosmological interest, such as the curvature perturbations and the Mukhanov-Sasaki variable, we report the gauge dependent curvature in terms of GLC perturbations. This will be useful for later evaluations and is given by
\begin{align}
    \psi+\frac{1}{3}\Delta_{3}E
 =&\frac{1}{\Delta_{3}}\left[ \frac{1}{2}\left(\frac{1}{r}\partial_{w}+\frac{1}{r^{2}}-\frac{1}{2r^{2}}D^{2}\right)N+\frac{1}{4r^{2}}\left(D^{2}\right)^{2}\mu+\frac{1}{2r^{2}}D^{2}\mu\right.\nonumber\\
 & \left.+\frac{1}{2}\left(\partial_{w}+\frac{3}{r}\right)D^{2}u-\left(\partial_{w}^{2}+\frac{3}{r}\partial_{w}+\frac{1}{r^{2}}+\frac{1}{2r^{2}}D^{2}\right)\nu\right] \,,
\label{eq:gaugedependentcurvature}
\end{align}
where we have explicitly used also the SPS decomposition introduced in Eq.~\eqref{eq:sphere-decomp}. As a consistency check, we notice that Eq.~\eqref{eq:gaugedependentcurvature} does not depend on pseudo-scalar perturbations. The physical reason for that stands in the fact that l.h.s. of Eq.~\eqref{eq:gaugedependentcurvature} is entirely sourced by scalar perturbations in the SVT decomposition and this means that no magnetic modes can be present on the r.h.s. as well. Since in the SPS decomposition only the pseudo-scalars can source the magnetic modes (see \cite{Mitsou:2020czr} for a detailed discussion in this regard), they cannot appear on the r.h.s. We will discuss in details this relation in Sect. \ref{GaugeInv} for the science case study of the tensor perturbations where both electric and magnetic modes are present.

\subsection{Gauge Invariant Quantities} \label{GaugeInv}

With the formalism introduced above, we can straightforwardly focus on some relevant gauge invariant quantities within our light-cone perturbation theory. By using Eq.~\eqref{eq:gaugedependentcurvature},
we have that the quantities defined in Eqs.~\eqref{eq:GI-quantities} are
\begin{align}
\zeta
=& -\frac{1}{3}\frac{\delta\rho}{\rho+p}+\frac{1}{\Delta_{3}}\left[ \frac{1}{2}\left(\frac{1}{r}\partial_{w}+\frac{1}{r^{2}}-\frac{1}{2r^{2}}D^{2}\right)N+\frac{1}{4r^{2}}\left(D^{2}\right)^{2}\mu+\frac{1}{2r^{2}}D^{2}\mu\right.\,\nonumber\\
&+ \left.\frac{1}{2}\left(\partial_{w}+\frac{3}{r}\right)D^{2}u-\left(\partial_{w}^{2}+\frac{3}{r}\partial_{w}+\frac{1}{r^{2}}+\frac{1}{2r^{2}}D^{2}\right)\nu\right]\,,\nonumber\\
\mathcal{R}
=& H\frac{\delta\varphi}{\partial_{\tau}\varphi}+\frac{1}{\Delta_{3}}\left[ \frac{1}{2}\left(\frac{1}{r}\partial_{w}+\frac{1}{r^{2}}-\frac{1}{2r^{2}}D^{2}\right)N+\frac{1}{4r^{2}}\left(D^{2}\right)^{2}\mu+\frac{1}{2r^{2}}D^{2}\mu\right.\,\nonumber\\
&+ \left.\frac{1}{2}\left(\partial_{w}+\frac{3}{r}\right)D^{2}u-\left(\partial_{w}^{2}+\frac{3}{r}\partial_{w}+\frac{1}{r^{2}}+\frac{1}{2r^{2}}D^{2}\right)\nu\right]\,,
\label{eq:GI_scalars}
\end{align}
and we recall that $Q\equiv\partial_\tau\delta\varphi/H$ and $H\equiv\pa_\tau a/a$.
We notice that all the quantities in Eqs.~\eqref{eq:GI_scalars} do not depend on perturbations which are null in the linearized GLC gauge, namely $V_a=L=0$ and $N+2aM=0$ (see Eq.~(3.1) of \cite{Fanizza:2020xtv}). This means that a direct computation of $\zeta$, $\mathcal{R}$ and $Q$ within the linearized GLC gauge is equivalent in form to the actual gauge invariant expressions for these quantities. Moreover, as a side fact, we checked that our expression for $\mathcal{R}$ agrees with the result of \cite{Frob:2021ore}, where the linearized GLC gauge has been fixed from the beginning. This agreement is shown in Appendix \ref{app:FL}.

Finally, we compute the tensor components as presented in Eq.~\eqref{eq:h_ij}
by using Eqs.~\eqref{eq:scal}, \eqref{eq:ninjcij} and \eqref{eq:gaugedependentcurvature}. We then get the result for the radial projection
\begin{align}
h_{rr}
 =&\, \frac{1}{2}N+\frac{1}{2\Delta_{3}}\partial_{w}^{2}\left(N+4\nu\right)-\frac{1}{\Delta_{3}}\partial_{w}\left[\left(\partial_{w}+\frac{2}{r}\right)N+D^{2}u-\frac{4}{r}\nu\right]
\nonumber\\
 &+ \frac{1}{\Delta_{3}}\left(1+\frac{\partial_{w}^{2}}{\Delta_{3}}\right)\left[ \frac{1}{2}\left(\frac{1}{r}\partial_{w}+\frac{1}{r^{2}}-\frac{1}{2r^{2}}D^{2}\right)N+\frac{1}{4r^{2}}\left(D^{2}\right)^{2}\mu+\frac{1}{2r^{2}}D^{2}\mu\right.\,\nonumber\\
&+ \left.\frac{1}{2}\left(\partial_{w}+\frac{3}{r}\right)D^{2}u-\left(\partial_{w}^{2}+\frac{3}{r}\partial_{w}+\frac{1}{r^{2}}+\frac{1}{2r^{2}}D^{2}\right)\nu\right]\,,
\label{eq:hrr-GLC}
\end{align}
for the radial-angular component
\begin{align}
h_{ar}
 =&\,\frac{r^{2}}{2}\left(D_{a}u+\tilde{D}_{a}\hat{u}\right)+\frac{1}{2\Delta_{3}}\left(\partial_{w}-\frac{1}{r}\right)D_{a}\left(N+4\nu\right)
 - \frac{D_{a}}{2\Delta_{3}}\left[\left(\partial_{w}+\frac{2}{r}\right)N+D^{2}u-\frac{4}{r}\nu\right]\,
\nonumber\\
 &-\frac{1}{2\Delta_{3}}\left(\partial_{w}-\frac{2}{r}\right)\left[\left(\partial_{w}+\frac{2}{r}\right)\left(r^{2}D_{a}u+r^{2}\tilde{D}_{a}\hat{u}\right)+2D_{a}\nu
 \right.\,\nonumber\\
&\left.
+D_{a}\left(D^{2}\mu+2\mu\right)+\tilde{D}_{a}\left(D^{2}\hat{\mu}+2\hat{\mu}\right)\right]\nonumber\\
&+  \frac{1}{\left(\Delta_{3}\right)^{2}}\left[\left(\partial_{w}-\frac{1}{r}\right)D_{a}\right]\left[ \frac{1}{2}\left(\frac{1}{r}\partial_{w}+\frac{1}{r^{2}}-\frac{1}{2r^{2}}D^{2}\right)N+\frac{1}{4r^{2}}\left(D^{2}\right)^{2}\mu+\frac{1}{2r^{2}}D^{2}\mu\right.\,
 \nonumber\\
&+\left.\frac{1}{2}\left(\partial_{w}+\frac{3}{r}\right)D^{2}u-\left(\partial_{w}^{2}+\frac{3}{r}\partial_{w}+\frac{1}{r^{2}}+\frac{1}{2r^{2}}D^{2}\right)\nu\right]\,,
\label{eq:har-GLC}
\end{align}
and for the pure angular part
\begin{align}
h_{ab}
=&\,r^{2}\left(\bar{q}_{ab}\nu+D_{ab}\mu+\tilde{D}_{ab}\hat{\mu}\right)+\frac{1}{\Delta_{3}}\left(D_{a}D_{b}+\frac{\gamma_{ab}}{r}\partial_{w}\right)\left[\frac{1}{2}\left(N+4\nu\right)\right]\,\nonumber\\
 &- \frac{1}{\Delta_{3}}\left\{ \frac{\gamma_{ab}}{r}\left[\left(\partial_{w}
+\frac{2}{r}\right)N+D^{2}u\right]\right\}-\frac{1}{\Delta_{3}}\left\{ \left(\partial_{w}+\frac{2}{r}\right)\left(r^{2}D_{a}D_{b}u+r^{2}\tilde{D}_{ab}\hat{u}\right)\,\right.\nonumber\\
&+\left.D_{a}D_{b}\left(2\nu+D^{2}\mu+2\mu\right)+\tilde{D}_{ab}\left(D^{2}\hat{\mu}+2\hat{\mu}\right)\right\} \nonumber\\
 &+\frac{1}{\Delta_{3}}\left[\gamma_{ab}+\frac{\left(D_{a}D_{b}+\frac{\gamma_{ab}}{r}\partial_{w}\right)}{\Delta_{3}}\right]\left\{ \frac{1}{2}\left(\frac{1}{r}\partial_{w}+\frac{1}{r^{2}}-\frac{1}{2r^{2}}D^{2}\right)N+\frac{1}{4r^{2}}\left(D^{2}\right)^{2}\mu\right.\, \nonumber\\
 &+\left.\frac{1}{2r^{2}}D^{2}\mu+\frac{1}{2}\left(\partial_{w}+\frac{3}{r}\right)D^{2}u-\left(\partial_{w}^{2}+\frac{3}{r}\partial_{w}+\frac{1}{r^{2}}+\frac{1}{2r^{2}}D^{2}\right)\nu\right\}\,,
\label{eq:hab-GLC}
\end{align}
which are the first order gauge invariant tensor perturbation within our new perturbation theory. In Appendix~\ref{app:GT}, we explicitly checked the expected gauge transformation of the SVT perturbations in terms of the light-cone perturbations. As expected, tensor perturbations are gauge invariant.

In Appendix \ref{app:FL}, we have also checked that our results for the tensor perturbations agrees with the ones which can be evaluated by following the procedure outlined in \cite{Frob:2021ore}, where the authors provide a general formula to evaluate the tensor perturbations, even though they do not provide their explicit evaluations.

At first look, the expressions for the light-cone perturbations look a way messier than the ones typically used in the SVT decomposition for the 1+3 spacetime foliation. Despite this complication, the advantage in using the light-cone perturbations stands in the fact that our expressions are ready to be decomposed in terms of electric and magnetic modes. We partially discussed this feature after Eq.~\eqref{eq:gaugedependentcurvature}. To show the actual advantage of this decomposition, we will introduce in the next Section the helicity basis. Even though tensor perturbations contain both scalar and pseudo-scalar contributions, let us note that the pure radial projection is exclusively sourced by scalar perturbations. This is not surprising, since $h_{rr}$ is just the helicity-0 mode of the tensor perturbations, hence it cannot contain any pseudo-scalar term.

\section{A coordinate independent Scalar-PseudoScalar decomposition}
\label{sec:gen}
The results obtained in the previous sections show that a link between the light-cone and the SVT perturbations does not necessarily bring to manageable expressions. It is worth then to provide a way to classify perturbations in the SPS manner which is coordinate independent, in order to better understand the link between different perturbative schemes. To this end, we first consider an unspecified metric $g_{\mu\nu}$ which can be linearized as follows
\begin{align}
g_{\mu\nu}\approx\bar{g}_{\mu\nu}+\delta g_{\mu\nu}\,,
\label{eq:linearized}
\end{align}
where $\bar{g}_{\mu\nu}$ is a generic background and $\delta g_{\mu\nu}$ are a set of metric perturbations added on top of the former. With respect to the background metric, we then define the following set of vectors
\begin{align}
\u^\mu \u_\mu=-1
\qquad,\qquad
\n^\mu \n_\mu =1\,,
\qquad,\qquad \n^\mu u_\mu=0\,.
\label{eq:vectors}
\end{align}
Here $\u^\mu$ plays the role of the time-like flow for the $\g_{\mu\nu}$, whereas $\n^\mu$ is the "radial" space-like direction within the spanned 3-D hyper-surface orthogonal to $\u^\mu$. Since this hyper-surface is 3-D, it admits three independent directions to form an orthonormal space-like basis. Indeed, besides $\n^\mu$, we can also introduce $\e^\mu_A$, where $A=1,2$ and
\begin{equation}
\e^\mu_A\e_{\mu B}=\delta_{AB}
\qquad\text{and}\qquad\n^\mu\e_{\mu A}=0\,.
\label{eq:8.3}
\end{equation}
We then identify $\{ \n^\mu,\e^\mu_A \}$ as the orthonormal space-like triad and $\e^\mu_A$ as the unit-vectors orthogonal to the "radial" direction $\bar{n}^\mu$. The vectors $\e^\mu_A$ can be rewritten in the so-called {\it helicity basis}, namely a complex linear combination of their elements given by
\begin{equation}
\e^\mu_\pm\equiv \frac{1}{\sqrt{2}}\left( \e^\mu_1\pm i \e^\mu_2 \right)\,,
\label{eq:hel}
\end{equation}
where it follows that $\e^\mu_\pm\e_{\mu \pm}=0$ and $\e^\mu_\pm\e_{\mu \mp}=1$. The advantage in working with Eq.~\eqref{eq:hel} stands in the fact that $\e^\mu_\pm$ have well-defined transformation rules under 2-D rotations in the Euclidean plane spanned by $\e^\mu_A$ itself. Indeed, such a rotation is given by
\begin{equation}
R^A_B=\cos\beta\,\delta^A_B+i\sin\beta\,\left(\sigma_2\right)^A_B\,,
\end{equation}
where $\sigma_2$ is the antisymmetric Pauli matrix and $\beta$ is the rotation angle. Hence, for the rotated basis $\widetilde{\e^\mu_A}\equiv R^B_A\e^\mu_B$, the helicity basis given by $\widetilde{\e^\mu_\pm}\equiv \frac{1}{\sqrt{2}}\left( \widetilde{\e^\mu_1}\pm i \widetilde{\e^\mu_2} \right)$ transforms as
\begin{equation}
\widetilde{\e^\mu_\pm}= e^{\pm i\beta}\,\e^\mu_\pm\,.
\end{equation}
In this way, the perturbations $\delta g_{\mu\nu}$ can be classified according to their transformation properties under such a rotation. Explicitly, we have 4 helicity-0 projections, i.e.
\begin{equation}
\S\equiv \u^\mu \u^\nu \delta g_{\mu\nu}
\quad,\quad
\T_\rVert\equiv \n^\mu \n^\nu \delta g_{\mu\nu}
\quad,\quad
\V_\rVert\equiv \n^\mu \u^\nu \delta g_{\mu\nu}
\quad,\quad
\T\equiv \e^\mu_+ \e^\nu_- \delta g_{\mu\nu}\,,
\label{eq:hel0}
\end{equation}
two helicity-1 projections
\begin{equation}
\V_\pm\equiv \u^\mu \e^\nu_\pm\delta g_{\mu\nu}
\quad,\quad
\T_{\rVert\pm}\equiv \n^\mu \e^\nu_\pm\delta g_{\mu\nu}\,,
\label{eq:hel1}
\end{equation}
and a single set of helicity-2 perturbations, namely
\begin{equation}
\T_{\pm\pm}\equiv \e^\mu_\pm \e^\nu_\pm \delta g_{\mu\nu}\,.
\label{eq:hel2}
\end{equation}

This procedure to classify linear perturbations has the manifest advantage that is independent of the (possible) symmetries present in the Killing vectors of $\g_{\mu\nu}$. This is not the case for the standard cosmological perturbation theory, where the SVT representation with respect to the rotational symmetry of the background is largely used. Moreover, also the gauge properties of our projections can be treated without an explicit notion of the adopted background. Indeed, under a linear diffeomorphism $x^\mu\rightarrow \tilde{x}^\mu=x^\mu+\epsilon^\mu$, the metric transforms as
\begin{equation}
\widetilde{\delta g}_{\mu\nu}(x^\alpha)=\delta g_{\mu\nu}(x^\alpha)-
\bar{\nabla}_\mu\epsilon_\nu(x^\alpha)
-\bar{\nabla}_\nu\epsilon_\mu(x^\alpha)\,,
\end{equation}
where $\bar\nabla_\mu$ is the covariant derivative with respect to $\bar{g}_{\mu\nu}$. In this way, we can project $\epsilon^\mu$ as well onto the background vectors and get
\begin{equation}
T\equiv \u^\mu\epsilon_\mu
\qquad,\qquad
R\equiv \n^\mu\epsilon_\mu
\qquad,\qquad
X_\pm\equiv \e^\mu_\pm \epsilon_\mu\,.
\label{eq:GM}
\end{equation}
Hence, it follows that helicity-0 perturbations \eqref{eq:hel0} transform as
\begin{align}
\widetilde{\S}=&\,\S
-2\,\u^\mu\bar{\nabla}_\mu T
+2\left(\u^\mu\bar{\nabla}_\mu \u^\nu\right)\epsilon_\nu\,,
\nonumber\\
\widetilde{\T}_\rVert=&\,\T_\rVert
-2\,\n^\mu \bar{\nabla}_\mu R
+2\left(\n^\mu \bar{\nabla}_\mu \n^\nu\right)\epsilon_\nu\,,
\nonumber\\
\widetilde{\V}_\rVert=&\,\V_\rVert
-\n^\mu\bar{\nabla}_\mu T
-\u^\mu\bar{\nabla}_\mu R
+\left(\n^\mu\bar{\nabla}_\mu\u^\nu\right)\epsilon_\nu
+\left(\u^\mu\bar{\nabla}_\mu\n^\nu\right)\epsilon_\nu\,,
\nonumber\\
\widetilde{\T}=&\,\T
-\e^\mu_+\bar{\nabla}_\mu X_-
-\e^\mu_-\bar{\nabla}_\mu X_+
+\left(\e^\mu_+\bar{\nabla}_\mu\e^\nu_-\right)\epsilon_\nu
+\left(\e^\mu_-\bar{\nabla}_\mu\e^\nu_+\right)\epsilon_\nu\,,
\label{eq:hel0_GT}
\end{align}
helicity-1 projections in \eqref{eq:hel1} change according to
\begin{align}
\widetilde{\V}_\pm=&\,\V_\pm
-\u^\mu\bar{\nabla}_\mu X_\pm
-\e^\mu_\pm\bar{\nabla}_\mu T
+\left(\e^\mu_\pm\bar{\nabla}_\mu\u^\nu\right)\epsilon_\nu
+\left(\u^\mu\bar{\nabla}_\mu\e^\nu_\pm\right)\epsilon_\nu\,,
\nonumber\\
\widetilde{\T}_{\rVert\pm}=&\,\T_{\rVert\pm}
-\n^\mu\bar{\nabla}_\mu X_\pm
-\e^\mu_\pm\bar{\nabla}_\mu R
+\left(\e^\mu_\pm\bar{\nabla}_\mu\n^\nu\right)\epsilon_\nu
+\left(\n^\mu\bar{\nabla}_\mu\e^\nu_\pm\right)\epsilon_\nu\,,
\label{eq:hel1_GT}
\end{align}
and, finally, helicity-2 perturbations \eqref{eq:hel2} transform as
\begin{equation}
\widetilde{\T}_{\pm\pm}=\,\T_{\pm\pm}
-2\,\e^\mu_\pm\bar{\nabla}_\mu X_\pm
+2\left(\e^\mu_\pm\bar{\nabla}_\mu\e^\nu_\pm\right)\epsilon_\nu\,.
\label{eq:hel2_GT}
\end{equation}

At this point, we first remark that the projected gauge modes \eqref{eq:GM} are scalars under background coordinate transformations, such that all the covariant derivatives $\bar\nabla_\mu$ acting on $T,\,R,\,X_\pm$ in \eqref{eq:hel0_GT}-\eqref{eq:hel2_GT} can be immediately written as ordinary partial derivatives. Moreover, $T$ and $R$ can be respectively interpreted as "time" and "radial" gauge modes, although the explicit form of time-like and space-like coordinate for $\g_{\mu\nu}$ has not been provided. The same happens for $X_\pm$, which can be read as the clockwise and anti-clockwise gauge modes on the 2-D space-like screen where the angular coordinates are typically defined.

To conclude this section, we do a further step in the gauge transformations \eqref{eq:hel0_GT}-\eqref{eq:hel2_GT}. Indeed, it is still a bit unsatisfactory to have those gauge transformation explicitly in terms of $\epsilon_\mu$ rather than its projections $T,\,R,\,X_\pm$. To this end, we then write explicitly each vector in the form $h^\mu\bar\nabla_\mu f^\nu$ in terms of its projections onto $\{ \u_\nu,\n_\nu,\e_{\nu\pm}\}$, namely
\begin{equation}
h^\mu\bar\nabla_\mu f^\nu
=\left(\u_\rho h^\mu\bar\nabla_\mu f^\rho\right) \u^\nu
+\left(\n_\rho h^\mu\bar\nabla_\mu f^\rho\right) \n^\nu
+\left(\e_{\rho\mp} h^\mu\bar\nabla_\mu f^\rho\right) \e^\nu_\pm\,,
\end{equation}
where the last component projects $\mp$ helicity onto the $\pm$ vector because of the properties $\e^\mu_{\pm}\e_{\mu\mp}=1$ and $\e^\mu_{\pm}\e_{\mu\pm}=0$.
In this way, the gauge transformation can be written as
\begin{align}
\widetilde{\S}=&\,\S
-2\,\u^\mu\bar{\nabla}_\mu T
+2\left(\u_\nu\u^\mu\bar{\nabla}_\mu \u^\nu\right) T
+2\left(\n_\nu\u^\mu\bar{\nabla}_\mu \u^\nu\right) R
+2\left(\e_{\nu\mp}\u^\mu\bar{\nabla}_\mu \u^\nu\right) X_\pm\,,
\nonumber\\
\nonumber\\
\widetilde{\T}_\rVert=&\,\T_\rVert
-2\,\n^\mu \bar{\nabla}_\mu R
+2\left(\u_\nu\n^\mu \bar{\nabla}_\mu \n^\nu\right) T
+2\left(\n_\nu\n^\mu \bar{\nabla}_\mu \n^\nu\right) R
+2\left(\e_{\nu\mp}\n^\mu\bar{\nabla}_\mu \n^\nu\right) X_\pm\,,
\nonumber\\
\nonumber\\
\widetilde{\V}_\rVert=&\,\V_\rVert
-\n^\mu\bar{\nabla}_\mu T
-\u^\mu\bar{\nabla}_\mu R
\nonumber\\
&+\left(\u_\nu\n^\mu\bar{\nabla}_\mu\u^\nu\right) T
+\left(\n_\nu\n^\mu\bar{\nabla}_\mu\u^\nu\right) R
+\left(\e_{\nu\mp}\n^\mu\bar{\nabla}_\mu\u^\nu\right) X_\pm
\nonumber\\
&+\left(\u_\nu\u^\mu\bar{\nabla}_\mu\n^\nu\right) T
+\left(\n_\nu\u^\mu\bar{\nabla}_\mu\n^\nu\right) R
+\left(\e_{\nu\mp}\u^\mu\bar{\nabla}_\mu\n^\nu\right) X_\pm\,,
\nonumber\\
\nonumber\\
\widetilde{\T}=&\,\T
-\e^\mu_+\bar{\nabla}_\mu X_-
-\e^\mu_-\bar{\nabla}_\mu X_+
\nonumber\\
&+\left(\u_\nu\e^\mu_+\bar{\nabla}_\mu\e^\nu_-\right) T
+\left(\n_\nu\e^\mu_+\bar{\nabla}_\mu\e^\nu_-\right) R
+\left(\e_{\nu\mp}\e^\mu_+\bar{\nabla}_\mu\e^\nu_-\right) X_\pm
\nonumber\\
&+\left(\u_\nu\e^\mu_-\bar{\nabla}_\mu\e^\nu_+\right) T
+\left(\n_\nu\e^\mu_-\bar{\nabla}_\mu\e^\nu_+\right) R
+\left(\e_{\nu\mp}\e^\mu_-\bar{\nabla}_\mu\e^\nu_+\right) X_\pm\,,
\label{eq:hel0_GT_p2}
\end{align}
for the helicity-0 components, as
\begin{align}
\widetilde{\V}_\pm=&\,\V_\pm
-\u^\mu\bar{\nabla}_\mu X_\pm
-\e^\mu_\pm\bar{\nabla}_\mu T
\nonumber\\
&+\left(\u_\nu\e^\mu_\pm\bar{\nabla}_\mu\u^\nu\right) T
+\left(\n_\nu\e^\mu_\pm\bar{\nabla}_\mu\u^\nu\right) R
+\left(\e_{\nu\mp}\e^\mu_\pm\bar{\nabla}_\mu\u^\nu\right) X_\pm
\nonumber\\
&+\left(\u_\nu\u^\mu\bar{\nabla}_\mu\e^\nu_\pm\right) T
+\left(\n_\nu\u^\mu\bar{\nabla}_\mu\e^\nu_\pm\right) R
+\left(\e_{\nu\mp}\u^\mu\bar{\nabla}_\mu\e^\nu_\pm\right) X_\pm\,,
\nonumber\\
\nonumber\\
\widetilde{\T}_{\rVert\pm}=&\,\T_{\rVert\pm}
-\n^\mu\bar{\nabla}_\mu X_\pm
-\e^\mu_\pm\bar{\nabla}_\mu R
\nonumber\\
&+\left(\u_\nu\e^\mu_\pm\bar{\nabla}_\mu\n^\nu\right) T
+\left(\n_\nu\e^\mu_\pm\bar{\nabla}_\mu\n^\nu\right) R
+\left(\e_{\nu\mp}\e^\mu_\pm\bar{\nabla}_\mu\n^\nu\right) X_\pm
\nonumber\\
&+\left(\u_\nu\n^\mu\bar{\nabla}_\mu\e^\nu_\pm\right) T
+\left(\n_\nu\n^\mu\bar{\nabla}_\mu\e^\nu_\pm\right) R
+\left(\e_{\nu\mp}\n^\mu\bar{\nabla}_\mu\e^\nu_\pm\right) X_\pm\,,
\label{eq:hel1_GT_p2}
\end{align}
for the helicity-1 projections and finally as
\begin{equation}
\widetilde{\T}_{\pm\pm}=\,\T_{\pm\pm}
-2\,\e^\mu_\pm\bar{\nabla}_\mu X_\pm
+2\left(\u_\nu\e^\mu_\pm\bar{\nabla}_\mu\e^\nu_\pm\right) T
+2\left(\n_\nu\e^\mu_\pm\bar{\nabla}_\mu\e^\nu_\pm\right) R
+2\left(\e_{\nu\mp}\e^\mu_\pm\bar{\nabla}_\mu\e^\nu_\pm\right) X_\pm\,,
\label{eq:hel2_GT_p2}
\end{equation}
for the helicity-2 ones. Although Eqs.~\eqref{eq:hel0_GT_p2}-\eqref{eq:hel2_GT_p2} might look more complicated than Eqs.~\eqref{eq:hel0_GT}-\eqref{eq:hel2_GT}, they have the advantage that all the gauge modes are explicitly written in terms of their scalar projections $T,\,R$ and $X_\pm$. This renders their manipulation quite easier when the background (and its symmetries, if any!) are left unspecified.

The fact that the projections \eqref{eq:hel0}-\eqref{eq:hel2} and their gauge transformations~\eqref{eq:hel0_GT_p2}-\eqref{eq:hel2_GT_p2} are readily given only in terms of scalar quantities means that they can be evaluated for different backgrounds and directly compared between each other.

We remark that the projections that we have built here rely on the possibility to define a reference background
for which the four vectors $\{\bar u^\mu,\bar n^\mu,\bar e^\mu_\pm\}$ are
defined. Once this splitting is obtained, the perturbations can be projected onto this
set of vectors, regardless of their perturbative order and their transformation properties
under a finite redefinition of the coordinate background are given at any
perturbative order and regardless of the background symmetries.

To conclude this section, we underline that for the gauge transformations~\eqref{eq:hel0_GT_p2}-\eqref{eq:hel2_GT_p2} no notion about the transport of the basis $\{ \u^\mu,\n^\nu,\e^\mu_\pm \}$ has been adopted. Particular choices can heavily simplify the overall form of the gauge transformations. This will be manifest in the subsequent sections.

\subsection{Geodesic Light-Cone background}

Here we want to apply the general formalism developed in the previous section to the perturbation theory on the light-cone already introduced in \cite{Fanizza:2020xtv}. We can now profit of the properties of the GLC gauge to find the suited background basis for the helicity projections presented in this section. Since $\bar{f}_{\mu\nu}$ shares the same features of $g^{GLC}_{\mu\nu}$, we start from the background vectors
\begin{equation}
\u_\mu=-\delta^\tau_\mu
\qquad\text{and}\qquad
\k_\mu=\delta^w_\mu\,,
\label{eq:uk}
\end{equation}
and build $\n_\nu$ as
\begin{equation}
\n_\nu=\u_\mu+\left( \u^\alpha\k_\alpha \right)^{-1}\k_\mu
=-\delta^\tau_\mu+a\,\delta^w_\mu\,.
\label{eq:n}
\end{equation}
Indeed, it is straightforward to check that $\u^\mu\u_\mu=-1$ and $\n^\mu\n_\mu=1$. For what concerns the basis $\e^\nu_A$, the requirement that $\u_\nu\e^\nu_A=\n_\nu\e^\nu_A=0$ returns that they can only admit non-null components in their angular part. We fix the latter as follows
\begin{equation}
\e_{\nu 1}=ar\,\delta^{\theta^1}_\nu
\qquad,\qquad
\e_{\nu 2}=ar\sin\theta^1\,\delta^{\theta^2}_\nu\,.
\label{eq:lin_basis}
\end{equation}
We remark that this choice is not only consistent with the requirements \eqref{eq:8.3} but also provides that
\begin{equation}
\Pi^\mu_\nu \k^\alpha \bar\nabla_\alpha \e^\nu_A=0\,,
\label{eq:Sachs_transport}
\end{equation}
where $\Pi_{\mu\nu}\equiv \bar{f}_{\mu\nu}+\u_\mu\u_\nu-\n_\mu\n_\nu$ is the 2-D sub-manifold orthogonal to $\u_\mu$ and $\n_\mu$. In other words, $\e^\mu_A$ is chosen to be the background Sachs basis \cite{Fanizza:2013doa} and a notion of transport for it is provided.

Thanks to Eqs.~\eqref{eq:lin_basis}, it is easy now to obtain also the circular basis \eqref{eq:hel} as
\begin{equation}
\e_{\nu \pm}=\frac{ar}{\sqrt{2}}\,\left(\delta^{\theta^1}_\nu\pm i \sin\theta^1\,\delta^{\theta^2}_\nu\right)\,.
\label{eq:circ_basis}
\end{equation}
With Eqs.~\eqref{eq:uk}, \eqref{eq:n} and \eqref{eq:circ_basis}, the SPS perturbations \eqref{eq:hel0}-\eqref{eq:hel2} in the light-cone framework are readily written as four helicity-0 perturbations
\begin{equation}
\S=a^2L+2aM+N
\quad,\quad
\T_\rVert= N
\quad,\quad
\V_\rVert=N+aM
\quad,\quad
\T=2\,\nu\,,
\label{eq:hel0_LC}
\end{equation}
two helicity-1 projections
\begin{equation}
\V_\pm= r\,\pa_\pm\left[\left(u+av\right)\mp i\left(\hat{u}+a\hat{v}\right)\right]
\quad,\quad
\T_{\rVert\pm}= r\,\pa_\pm \left(u\mp i\hat{u}\right)\,,
\label{eq:hel1_LC}
\end{equation}
and a single set of helicity-2 perturbations, namely
\begin{equation}
\T_{\pm\pm}=\, \pa^2_\pm\left(\mu\mp i\hat{\mu}\right)\,.
\label{eq:hel2_LC}
\end{equation}

Moreover, the background basis $\{\u_\mu,\n_\nu,\e_{\nu\pm}\}$ explicitly given by Eqs.~\eqref{eq:uk}, \eqref{eq:n} and \eqref{eq:circ_basis} also provides the needed transport properties of the basis along the manifold. These are resumed as follows
\begin{equation}
\u^\mu\bar\nabla_\mu \u^\nu=\u^\mu\bar\nabla_\mu \n^\nu=\u^\mu\bar\nabla_\mu \e^\nu_\pm=\n^\mu\bar\nabla_\mu \e^\nu_\pm=0\,,
\label{eq:3.27}
\end{equation}
where the transport of $\u^\mu$ is a direct consequence of the fact that $\u^\mu$ follows a time-like geodesic flow and the transport of $\e^\mu_\pm$ along $\u^\nu$ and $\n^\nu$ follows from Eq.~\eqref{eq:Sachs_transport}. In addition to them, we also have that
\begin{equation}
\n^\mu\bar\nabla_\mu \n^\rho=H\,\u^\rho
\qquad,\qquad
\n^\mu\bar\nabla_\mu \u^\rho=H\,\n^\rho\,,
\label{eq:3.28}
\end{equation}
and
\begin{equation}
\e^\mu_\pm\bar\nabla_\mu\u^\rho=H\e^\rho_\pm
\qquad,\qquad
\e^\mu_\pm\bar\nabla_\mu\n^\rho=\frac{1}{ar}\,\e^\rho_\pm\,.
\label{eq:3.29}
\end{equation}
Finally, for the circular basis, we have
\begin{equation}
\e^\mu_\pm\bar\nabla_\mu\e^\rho_\mp=H \u^\rho
-\frac{\n^\rho}{ar}
-\frac{\cot\theta}{\sqrt{2}\,ar}\e^\rho_\mp
\qquad\text{and}\qquad
\e^\mu_\pm\bar\nabla_\mu\e^\rho_\pm=\frac{\cot\theta}{\sqrt{2}\,ar}\e^\rho_\pm\,.
\label{eq:3.30}
\end{equation}

The knowledge of transport for the background basis then immediately returns (and simplifies) the gauge transformations of the helicity projections reported in \eqref{eq:hel0_GT}-\eqref{eq:hel2_GT}, namely
\begin{align}
\widetilde{\S}=&\,\S
-2\,\u^\mu\bar{\nabla}_\mu T\,,
\nonumber\\
\widetilde{\T}_\rVert=&\,\T_\rVert
-2\,\n^\mu \bar{\nabla}_\mu R
-2\,H T\,,
\nonumber\\
\widetilde{\V}_\rVert=&\,\V_\rVert
-\n^\mu\bar{\nabla}_\mu T-\u^\mu\bar{\nabla}_\mu R+HR\,,
\nonumber\\
\widetilde{\T}=&\,\T
-\e^\mu_+\bar{\nabla}_\mu X_-
-\e^\mu_-\bar{\nabla}_\mu X_+
-2\,H T
-\frac{2}{ar}R
-\frac{\cot\theta}{\sqrt{2}\,ar}X_-
-\frac{\cot\theta}{\sqrt{2}\,ar}X_+
\nonumber\\
=&\,\T
+\frac{1}{ar}\left( \pa_+X_-+\pa_-X_+\right)
-2\,H T
-\frac{2}{ar}R\,,
\label{eq:hel0_GT_p3}
\end{align}
for the helicity-0 perturbations, then
\begin{align}
\widetilde{\V}_\pm=&\,\V_\pm
-\u^\mu\bar{\nabla}_\mu X_\pm
-\e^\mu_\pm\bar{\nabla}_\mu T
+H X_\pm
=\,\V_\pm
-\u^\mu\bar{\nabla}_\mu X_\pm
-\frac{1}{ar} \pa_\pm T
+H X_\pm\,,
\nonumber\\
\widetilde{\T}_{\rVert\pm}=&\,\T_{\rVert\pm}
-\n^\mu\bar{\nabla}_\mu X_\pm
-\e^\mu_\pm\bar{\nabla}_\mu R
+\frac{1}{ar}X_\pm
=\,\T_{\rVert\pm}
-\n^\mu\bar{\nabla}_\mu X_\pm
-\frac{1}{ar} \pa_\pm R
+\frac{1}{ar}X_\pm\,,
\label{eq:hel1_GT_p3}
\end{align}
for the helicity-1 projections and finally
\begin{equation}
\widetilde{\T}_{\pm\pm}=\,\T_{\pm\pm}
-2\,\left(\e^\mu_\pm\bar{\nabla}_\mu X_\pm
-\frac{\cot\theta}{\sqrt{2}\,ar}X_\pm\right)
=\,\T_{\pm\pm}
-\frac{2}{ar}\pa_\pm X_\pm\,,
\label{eq:hel2_GT_p3}
\end{equation}
for the helicity-2 modes. We notice few points at this stage:
\begin{itemize}
\item in the last of Eqs~\eqref{eq:hel0_GT_p3} and \eqref{eq:hel2_GT_p3}, the emergent structure from general gauge transformations automatically reproduces the derivatives $\pa_\pm$, which are related to the spin-lowering and raising operators $\ds$ and $\bds$, discussed in Appendix \ref{app:swsh};
\item moreover, all the projections correctly transform according to their helicity properties. This is evident from the fact that $T$ and $R$ and helicity-0 gauge modes. Moreover, $X_\pm$ have helicity $\pm 1$ and the action of $\pa_\pm$ is such that it increases/decreases by one unity the helicity of the field they act on;
\item gauge transformations in Eqs.~\eqref{eq:hel0_GT_p3}-\eqref{eq:hel2_GT_p3} have the advantage to make manifest the physical interpretations of the gauge modes themselves, since $T$ can be interpreted as the time-like mode, $R$ is the radial mode and $X_\pm$ correspond to the angular gauge freedom. Clearly, it is possible to express them according to the infinitesimal shift $\xi^\mu$ of the light-cone coordinates $\left(\tau,w,\theta^a\right)\rightarrow\left(\tau,w,\theta^a\right)+\xi^\mu$, as done in \cite{Fanizza:2020xtv}, having then
\begin{equation}
T=-\xi^0
\quad,\quad
R=-\xi^0+a\xi^w
\quad,\quad
X_\pm=\e_{a\pm}\xi^a\,.
\end{equation}
It is evident, however, how this would mix the different gauge modes. Moreover, the expressions in terms of $T,\,R,\,X_\pm$ have the advantage to deal with scalar quantities. This is not the case if one works with $\xi^\mu$ instead.
\end{itemize}

\subsection{Friedmann-Lema\^itre-Robertson-Walker background}

We now want to apply our formalism for the projected inhomogeneities to the standard linear perturbation theory built on the FLRW background. In fact, our final goal is to connect the light-cone perturbations in the helicity basis to the standard perturbations in the SVT decomposition. To this end, we will profit of the invariance for finite background coordinate transformation of our perturbative scheme. In fact, by using the background coordinate transformation in Eqs.~\eqref{eq:2.18}, we get that for the FLRW backgound, our basis is
\begin{equation}
\u_\mu=-a\,\delta^\eta_\mu
\qquad,\qquad
\n_\mu=a\,\delta_\mu^r\,,
\qquad,\qquad
\e_{\nu 1}=ar\,\delta^{\theta^1}_\nu
\qquad,\qquad
\e_{\nu 2}=ar\sin\theta^1\,\delta^{\theta^2}_\nu\,.
\label{eq:335}
\end{equation}
Moreover, the transport equations are exactly the same as Eqs.~\eqref{eq:3.27}-\eqref{eq:3.30}, with the substitution $H=\Hcal/a$ and $r=w-\eta$.
In this way, we can immediately write the helicity-0 projections as
\begin{equation}
\S=-2\phi
\quad,\quad
\T_\rVert= \C_{rr}
\quad,\quad
\V_\rVert=-\B_r
\quad,\quad
\T=\frac{\bar\gamma^{ab}\C_{ab}}{2}\,.
\label{eq:hel0_FRW}
\end{equation}
The two helicity-1 projections become
\begin{equation}
\V_\pm=-a\,\e^a_\pm\B_a
\quad,\quad
\T_{\rVert\pm}=a\,\e^a_\pm\C_{ra}\,.
\label{eq:hel1_FRW}
\end{equation}
Finally the helicity-2 perturbations is given by
\begin{equation}
\T_{\pm\pm}=a^2\,\e^a_\pm\e^b_\pm\C_{ab}\,.
\label{eq:hel2_FRW}
\end{equation}
The knowledge of transport for the background basis then immediately returns (and simplify) the gauge transformations of the helicity projections. These are formally equivalent to \eqref{eq:hel0_GT_p2}-\eqref{eq:hel2_GT_p2}, once it is taken into account that $H=\Hcal/a$ and $r=w-\eta$ and now $T,\,R$ and $X_\pm$ are given by
\begin{equation}
T=-a\,\epsilon^\eta
\quad,\quad
R=a\,\epsilon^r
\quad,\quad
X_\pm=\e_{a\pm}\epsilon^a\,.
\label{eq:339}
\end{equation}

\section{{\it E} and {\it B} modes for tensor perturbations}
\label{sec:EB}
The results achieved in the previous sections allow to relate directly the projected perturbations as evaluated w.r.t. different backgrounds. Indeed, since the helicity perturbations are independent of the chosen background, we can straightforwardly equal Eqs.~\eqref{eq:hel0_FRW}-\eqref{eq:hel2_FRW} with \eqref{eq:hel0_LC}-\eqref{eq:hel2_LC}. We then obtain for the helicity-0 projections
\begin{equation}
\phi=-\frac{1}{2}\left(a^2L+2aM+N\right)
\quad,\quad
\C_{rr}=N
\quad,\quad
\B_r=-N-aM
\quad,\quad
\bar\gamma^{ab}\C_{ab}= 4\nu\,,
\label{eq:hel0_link}
\end{equation}
for the helicity-1
\begin{align}
\B_\pm\equiv&\,\e^a_\pm\B_a=-\frac{r}{a}\,\pa_\pm\left[\left(u+av\right)\mp i\left(\hat{u}+a\hat{v}\right)\right]\,,
\nonumber\\
\C_{r\pm}\equiv&\,\e^a_\pm\C_{ra}=\frac{r}{a}\,\pa_\pm \left(u\mp i\hat{u}\right)\,,
\label{eq:hel1_link}
\end{align}
and finally for the helicity-2
\begin{equation}
\C_{\pm\pm}\equiv\e^a_\pm\e^b_\pm\C_{ab}=\frac{2}{a^2}\,\pa^2_\pm\left(\mu\mp i\hat{\mu}\right)\,.
\label{eq:hel2_link}
\end{equation}
As a check, we notice that the results of Eqs.~\eqref{eq:hel0_link} are in agreement with the ones provided in \eqref{eq:Crr}.

A remarkable feature is that Eqs.~\eqref{eq:hel0_link}-\eqref{eq:hel2_link} are given by construction in terms of the spin raising and lowering operators. This is a clear advantage if one is interested in studying the $E$- and $B$- modes of the perturbations. Indeed, for a generic tensor field of rank $s$ $T_{{a_1}...{a_s}}$, we recall that the projections onto $\e^a_\pm$ can be written as
\begin{equation}
T_\pm\equiv \e^{a_1}_\pm...\e^{a_s}_\pm T_{{a_1}...{a_s}}\,,
\label{eq:proj}
\end{equation}
and then one can extract its $E$- and $B$- modes thanks to the action of the spin raising and lowering operators, respectively $\ds$ and $\bds$, as
\begin{align}
T^{\mathbf{E}}\equiv\frac{\bds^{\,s} T_++\ds^{\,s} T_-}{2}
\qquad\text{and}\qquad
T^{\mathbf{B}}\equiv -i\frac{\bds^{\,s} T_+-\ds^{\,s} T_-}{2}\,.
\label{eq:45}
\end{align}
We redirect to Appendix \ref{app:swsh} for details and useful relations about $\bds$ and $\ds$.

Let us notice that the projections \eqref{eq:proj} have helicity $s$. Hence, for the helicity-1 and 2 projections, we can directly obtain the $E$- and $B$- modes of the perturbations \eqref{eq:hel1_link} and \eqref{eq:hel2_link} as
\begin{align}
\mathcal{B}^{\mathbf{E}} \equiv&\,\frac{\bds\B_++\ds\B_-}{2}=\frac{r}{a\sqrt{2}}D^2 \left(u+av\right)
\quad,\quad
\mathcal{B}^{\mathbf{B}} \equiv\,-i\frac{\bds\B_+-\ds\B_-}{2}=\frac{r}{a\sqrt{2}}D^2 \left(\hat{u}+a\hat{v}\right)\,,\nonumber\\
\mathcal{C}_r^{\mathbf{E}} \equiv&\,\frac{\bds\C_{r+}+\ds\C_{r-}}{2}=-\frac{r}{a\sqrt{2}}D^2 u
\quad,\quad
\mathcal{C}_r^{\mathbf{B}} \equiv -i\frac{\bds\C_{r+}-\ds\C_{r-}}{2}=-\frac{r}{a\sqrt{2}}D^2 \hat{u}\,,\nonumber\\
\mathcal{C}^{\mathbf{E}}\equiv&\,\frac{\bds^2\C_{++}+\ds^2\C_{--}}{2}=\,\frac{\left(D^2+2\right)D^2\mu}{a^{2}}
\quad,\quad
\mathcal{C}^{\mathbf{B}} \equiv -i\frac{\bds^2\C_{++}-\ds^2\C_{--}}{2}=\frac{\left(D^2+2\right)D^2\hat\mu}{a^{2}}\,.\nonumber\\
\label{eq:E-B-modes}
\end{align}
In Eqs.~\eqref{eq:E-B-modes} we have widely made use of the properties reported in Appendix \ref{app:swsh} and used the relations $\partial_+\equiv -\ds/\sqrt{2}$ and $\partial_-\equiv -\bds/\sqrt{2}$.

We notice that an important connection emerges when we relate standard perturbations (on top of a 1+3 background) with the light-cone ones (on top of a 1+1+2 background). Indeed, as can be seen in Table \ref{tab:correspondence}, an interesting block correspondence between the two sets of perturbations arises. In particular, among the four helicity-0 projections, the trace of the angular part is equivalent to $\nu$ and decouples from the other projections. Hence, the pure time-time standard perturbation $\phi$ and the radial projections $\mathcal{B}_r$ and $\mathcal{C}_{rr}$ are entirely sourced to $N$, $M$ and $L$.
\begin{table}[ht!]
\centering
\begin{tabular}{|c|c||cccc|cccc|cc|}
\cline{3-12}
\multicolumn{1}{l}{}&&$\nu$&$N$&$M$&$L$&$u$&$v$&$\hat{u}$&$\hat{v}$&$\mu$&$\hat{\mu}$\\
\hline\hline
\multirow{4}{*}{Helicity~0}&$\mathcal{C}$&\multicolumn{1}{l;{1pt/1pt}}{\Vc}&&&&&&&\multicolumn{1}{l}{}&&\\ 
\cdashline{3-6}[1pt/1pt]&$\mathcal{C}_{rr}$&\multicolumn{1}{l;{1pt/1pt}}{}&\Vc&&\multicolumn{1}{l|}{}&&&&\multicolumn{1}{l}{}&&\\
&$\mathcal{B}_r$&\multicolumn{1}{l;{1pt/1pt}}{}&\Vc&\Vc&\multicolumn{1}{l|}{}&&&&\multicolumn{1}{l}{}&&\\
&$\phi$&\multicolumn{1}{l;{1pt/1pt}}{}&\Vc&\Vc&\multicolumn{1}{l|}{\Vc}&&&&\multicolumn{1}{l}{}&&\\ 
\cline{1-10}\multirow{4}{*}{Helicity~1}&$\mathcal{C}_r^\mathbf{E}$&&&&\multicolumn{1}{l|}{}&\Vc&\multicolumn{1}{l;{1pt/1pt}}{}&&&&\\
&$\mathcal{B}^\mathbf{E}$&&&&\multicolumn{1}{l|}{}&\Vc&\multicolumn{1}{l;{1pt/1pt}}{\Vc}&&&&\\ 
\cdashline{7-10}[1pt/1pt]&$\mathcal{C}_r^\mathbf{B}$&&&&&&\multicolumn{1}{l;{1pt/1pt}}{}&\Vc&\multicolumn{1}{l|}{}&&\\
&$\mathcal{B}^\mathbf{B}$&&&&&&\multicolumn{1}{l;{1pt/1pt}}{}&\Vc&\multicolumn{1}{l|}{\Vc}&&\\ 
\cline{1-2}\cline{7-12}\multirow{2}{*}{Helicity~2}&$\mathcal{C}^\mathbf{E}$&&&&\multicolumn{1}{l}{}&&&&\multicolumn{1}{l|}{}&\multicolumn{1}{l;{1pt/1pt}}{\Vc}&\\
\cdashline{11-12}[1pt/1pt]&$\mathcal{C}^\mathbf{B}$&&&&\multicolumn{1}{l}{}&&&&\multicolumn{1}{l|}{}&\multicolumn{1}{l;{1pt/1pt}}{}&\Vc\\
\hline
\end{tabular}
\caption{Block scheme for the correspondence between the 1+3 perturbations and the 1+1+2 ones, according to the explicit relations \eqref{eq:hel0_link}-\eqref{eq:hel2_link}.}
\label{tab:correspondence}
\end{table}
Instead, for what regards the helicity-1 standard perturbations, i.e. $\mathcal{B}_a$ and $\mathcal{C}_{ra}$, their $E$- and $B$- modes are respectively sourced by $u$, $v$ and $\hat{u}$, $\hat{v}$, whereas $\mu$ and $\hat{\mu}$ are in a one-to-one correspondence with the $E$- and $B$- modes of the helicity-2 perturbations.

We remark that, within the GLC gauge fixing, where $v=\hat{v}=L=0$ and $N=-2aM$, Eqs.~\eqref{eq:hel0_link}-\eqref{eq:hel2_link} become
\begin{align}
\phi=&0
\qquad,\qquad
\mathcal{C}_{rr}  =-2\,\mathcal{B}_{r}=N
\qquad,\qquad
\mathcal{T}=2\nu\,,\nonumber\\
\mathcal{B}^{\mathbf{E}}
=&-\mathcal{C}_r^{\mathbf{E}}=\frac{r}{a\sqrt{2}}D^2u
\qquad,\qquad
\mathcal{B}^{\mathbf{B}} =-\mathcal{C}_r^{\mathbf{B}}=\frac{r}{a\sqrt{2}}D^2\hat{u}\,,
\nonumber\\
\mathcal{C}^{\mathbf{E}}=&\,\frac{1}{a^{2}}\left[\left(D^2\right)^2\mu+2D^2\mu\right]
\qquad,\qquad
\mathcal{C}^{\mathbf{B}} =\frac{1}{a^{2}}\left[\left(D^2\right)^2\hat\mu+2D^2\hat\mu\right]\,.
\label{eq:GLC_E-B-modes}
\end{align}

Although the resume in Table \ref{tab:correspondence} shows a clear correspondence between the SPS perturbations in 1+3 and 1+1+2 frameworks, we have to do a further step to show how the expressions for $E$- and $B$- modes of tensor perturbations look like in the light-cone setup. Therefore, we conclude this section by discussing these explicit expressions as functions of the SPS light-cone perturbations, while the full derivation is reported in Appendix \ref{app:EB}.

Let us start from Eq.~\eqref{eq:BC} and project the tensor perturbations on the helicity basis $\e^i_\pm$
\begin{equation}
h_{\pm\pm}\equiv\bar{e}_{\pm}^{i}\bar{e}_{\pm}^{j}h_{ij}=  \frac{\C_{\pm\pm}}{2}-\bar{e}_{\pm}^{i}\bar{e}_{\pm}^{j}\nabla_{i}S_{j}= \frac{\C_{\pm\pm}}{2}-\frac{1}{ar}\partial_{\pm}S_{\pm}\,,
\label{eq:tensor-projections}
\end{equation}
where we have defined $S_{i}\equiv\partial_{i}E+F_{i}$ and used the results of Appendix \ref{app:swsh}. Using Eq.~\eqref{eq:hel2_link} we can write 
\begin{equation}
h_{\pm\pm}=\frac{1}{a^{2}}\partial_{\pm}^{2}\left(\mu\mp i\hat{\mu}\right)-\frac{1}{ar}\partial_{\pm}S{}_{\pm}\,.\label{eq:sp-2-tensor}
\end{equation}
Hence, thanks to Eqs.~\eqref{eq:45} the $E$- and $B$- modes are given by
\begin{equation}
h^{\mathbf{E}}= \left(D^{2}+2\right)\left(\frac{D^{2}\mu}{2a^{2}}+\frac{1}{ar\sqrt{2}}S^{\mathbf{E}}\right)
\qquad\text{and}\qquad
h^{\mathbf{B}}=  \left(D^{2}+2\right)\left(\frac{D^{2}\hat{\mu}}{2a^{2}}+\frac{1}{ar\sqrt{2}}S^{\mathbf{B}}\right)\,.\label{eq:E-B-tensor}
\end{equation}
As one can check from Eqs.~\eqref{eq:gaugetransf} and \eqref{eq:gaugetransfvector}, $S_i$ transforms under a gauge redefinition of the fields as $\tilde{S}_i=S_i-\epsilon_i$ and then, thanks to Eq.~\eqref{eq:339}, we have that $\tilde{S}_\pm=\S_\pm-X_\pm$ or, equivalently, in terms of the $E/B$ modes,
\begin{equation}
\tilde{S}^{\mathbf{E}}= S^{\mathbf{E}}-X^{\mathbf{E}}
\qquad\text{and}\qquad
\tilde{S}^{\mathbf{B}}= S^{\mathbf{B}}-X^{\mathbf{B}}\,,\label{eq:E-B-S-transform}
\end{equation}
where $X^{\mathbf{E}/\mathbf{B}}$ are given in the SPS decomposition by
\begin{equation}
X^{\mathbf{E}}=  -\frac{r}{a\sqrt{2}}D^{2}\chi
\qquad\text{and}\qquad
X^{\mathbf{B}}=  -\frac{r}{a\sqrt{2}}D^{2}\hat{\chi}\,.\label{eq:X-mode}
\end{equation}
By recalling that $\mu$ and $\hat\mu$ transform under a gauge redefinition of the fields as $\tilde{\mu}= \mu-\chi$ and $\tilde{\hat{\mu}}= \hat{\mu}-\hat{\chi}$, Eqs.~\eqref{eq:X-mode} explicitly prove that Eqs.~\eqref{eq:E-B-tensor} are gauge invariant. Finally, we have that the $E$- modes of the tensor perturbations are given by
\begin{align}
h^{\mathbf{E}}= & \frac{\left(D^{2}+2\right)}{2a^{2}}\left\{ D^{2}\mu-\frac{1}{r\Delta_{3}}\left[\left(\partial_{w}+\frac{3}{r}\right)\left(rD^{2}u\right)+\frac{\left(D^{2}+2\right)}{r}D^{2}\mu+\frac{2D^{2}}{r}\nu\right]\right\} \nonumber \\
 & +\frac{\left(D^{2}+2\right)}{2a^{2}r}\frac{1}{\Delta_{3}}\left\{ \frac{D^{2}}{r}\left\{ \frac{1}{3}\left(-\frac{2}{r}\partial_{w}+\Delta_{3}\right)\left\{ \frac{1}{\Delta^2_{3}}\left[\frac{1}{2}\left(\partial_{w}^{2}+\frac{5}{r}\partial_{w}+\frac{3}{r^{2}}\right)\left(N-2\nu\right)\right.\right.\right.\right.\nonumber \\
 & \left.\left.-\frac{1}{4\,r^{2}}D^{2}\left(N+4\nu\right)+\frac{1}{6\,r^{2}}\left(\partial_{w}+\frac{1}{r}\right)D^{2}u+\frac{3}{2\,r^{2}}D^{2}\nu
 +\frac{3}{4r^2}\left(D^2+2\right)D^2\mu\right]\right\} \nonumber \\
 & \left.\left.+\frac{2}{r}\frac{1}{\Delta_{3}}\left[\left(\frac{1}{3}\partial_{w}+\frac{1}{r}\right)\left(2N-4\nu\right)+D^{2}u\right]-\frac{1}{3}\left(N+4\nu\right)\right\} \right\}\,,
 \label{eq:hE}
\end{align}
whereas the $B$- modes are given by
\begin{align}
h^{\mathbf{B}}= & \frac{\left(D^{2}+2\right)}{2a^2}\left\{ D^{2}\hat{\mu}-\frac{1}{r\Delta_{3}}\left[\left(\partial_w+\frac{3}{r}\right)\left(rD^{2}\hat{u}\right)+\frac{\left(D^{2}+2\right)}{r}D^{2}\hat{\mu}\right]\right\} \,,
\label{eq:hB}
\end{align}

Eqs.~\eqref{eq:hE} and \eqref{eq:hB} provide the expressions of the two independent helicity-2 degrees of freedom in terms of the light-cone perturbations. Even though this result has been achieved without fixing any gauge, their values are given only in terms of the GLC gauge perturbations. This supports the claim that the non-linear GLC line element is a natural choice for the study of cosmological observables. As expected, we have that the electric modes are only sourced by scalar perturbations in the SPS scheme, whereas the magnetic modes are sourced by the pseudoscalar fields. Therefore, the importance of Eqs.~\eqref{eq:hE} and \eqref{eq:hB} stands in the fact that they explicitly provide the SPS fields combination to be identified with the radiative gravitational degrees of freedom.

Now, we can then use Eqs.~\eqref{eq:hel0_FRW}-\eqref{eq:hel2_FRW} and \eqref{eq:E-B-modes} to write the $E/B$ modes of gravitational waves in terms of the coordinate independent decomposition introduced in Sect.~\ref{sec:gen}. To this aim, we split
\begin{equation}
h^{\mathbf{E/B}}\equiv h^{\mathbf{E/B}}_2 + h^\mathbf{E}_{01}\,,
\end{equation}
where
\begin{align}
 h^{\mathbf{E/B}}_2\equiv&
 \frac{1}{2a^2}\left\{ \T^{\mathbf{E/B}}-\frac{\left(D^{2}+2\right)}{ar\Delta_{3}}\left[\left(\partial_{\rVert}+\frac{3}{r}\right)\T_{\rVert}^{\mathbf{E/B}}-\frac{1}{r\sqrt{2}}\T^{\mathbf{E/B}}\right]\right\}\,,
 \nonumber\\
 h^\mathbf{E}_{01}\equiv&
\frac{1}{2a^2}\left\{\frac{\left(D^{2}+2\right)}{ar\Delta_{3}}\left[\frac{1}{r\sqrt{2}}D^{2}\T
+\frac{D^{2}}{ar\sqrt{2}}\left[2\left(\psi+\frac{1}{3}\Delta_{3}E\right)-\Delta_{3}E
+\frac{2}{r}\left(\partial_{r}E+F_{r}\right)\right]\right]\right\}\,.
\nonumber\\
\label{eq:411}
\end{align}
The rationale behind this splitting stands in the fact that $h^{\mathbf{E/B}}_2$ is only sourced by $\mu,\,u$ and $\hat\mu,\,\hat{u}$ respectively for the electric and magnetic modes. This then reflects the symmetry in the SPS decomposition of $U_a$ and $\gamma_{ab}$. 
We remark that Eq.~\eqref{eq:411} is valid for both $E$- and $B$- modes but the second line is present only for the $E$- modes of the tensor perturbations. Furthermore, Eq.~\eqref{eq:411} allows us to interpret $h^{\mathbf{E}/\mathbf{B}}_2$ in a coordinate independent way as the actual content related to the gravitational wave propagation along the observed light-cone.

\subsection{Link with the initial conditions}
\label{sec:initial}

The results presented in the previous sections tell us what are the relevant combinations to be studied along the past light-cone. However, a neat connection with the primordial initial conditions is yet to be provided. In fact, usually initial conditions are given on spatial hypersurfaces at the horizon re-entering redshift $z_*$, whereas we have discussed how to study the radiative degrees of freedom along the light-cone.

In order to provide the initial conditions to be evolved, we refer to the results of \cite{Fanizza:2022wob}. In particular, the link between the $E/B$ modes found in Sect.~\ref{sec:EB} and the spectra for primordial tensor perturbations in terms of their linear polarization $h_{\pm 1}$ is given, according to Eq.~(B.14) of \cite{Fanizza:2022wob} as\footnote{To directly apply the results of \cite{Fanizza:2022wob} in this paper, the following dictionary is intended: $h^{\mathbf{E}/\mathbf{B}}=\sum_{\ell,m}\mathcal{T}^{E/B}_{\ell m}$}
\begin{align}
h^\mathbf{E}(z_*,\bf{n})
=&-4\pi\int\frac{d^3{\bf k}}{(2\pi)^3}
\sum_{\ell=0}^\infty\sum_{m=-\ell}^\ell\,\left( -i \right)^\ell
\left[\frac{\ell(\ell -1)}{2} \frac{j_{\ell }(kr_*)}{(kr_*)^2}
-j_{\ell }(kr_*)+\frac{j_{\ell -1}(kr_*)}{kr_*}\right]
\nonumber\\
=&-2\pi\int\frac{d^3{\bf k}}{(2\pi)^3}
\sum_{\ell=0}^\infty\,
\mathcal{K}^\mathbf{E}_{\ell}(k r_*)
\left[\sum_{m=-\ell}^\ell\left(\,_{-2}Y^*_{\ell m}(\hat{{\bf k}};{\bf E})+\,_{2}Y^*_{\ell m}(\hat{{\bf k}};{\bf E})\right)\,h_{+}(z_*,{\bf k})\right.
\nonumber\\
&\left.
-\sum_{m=-\ell}^\ell\left(\,_{-2}Y^*_{\ell m}(\hat{{\bf k}};{\bf E})
-\,_{2}Y^*_{\ell m}(\hat{{\bf k}};{\bf E})\right)\,h_{\times}(z_*,{\bf k})\right]\,,
\nonumber\\
h^\mathbf{B}(z_*,\bf{n})
=&-4\pi\int\frac{d^3{\bf k}}{(2\pi)^3}\sum_{\ell=0}^\infty\sum_{m=-\ell}^\ell\,\left( -i \right)^\ell
\left[(\ell +2)\frac{ j_{\ell}(kr_*)}{kr_*}-j_{\ell +1}(kr_*)\right]
\nonumber\\
&\times\sum_{p=\pm 1}p\,_{-2p}Y^*_{\ell m}(\hat{{\bf k}};{\bf E})\,h_p(z_*,{\bf k})
\nonumber\\
=&-2\pi\int\frac{d^3{\bf k}}{(2\pi)^3}
\sum_{\ell=0}^\infty\,
\mathcal{K}^\mathbf{B}_{\ell}(k r_*)
\left[\sum_{m=-\ell}^\ell\left(\,_{-2}Y^*_{\ell m}(\hat{{\bf k}};{\bf E})-\,_{2}Y^*_{\ell m}(\hat{{\bf k}};{\bf E})\right)\,h_{+}(z_*,{\bf k})\right.
\nonumber\\
&\left.-i\sum_{m=-\ell}^\ell\left(\,_{-2}Y^*_{\ell m}(\hat{{\bf k}};{\bf E})
+\,_{2}Y^*_{\ell m}(\hat{{\bf k}};{\bf E})\right)\,h_{\times}(z_*,{\bf k})\right]\,,
\label{eq:initial_conditions}
\end{align}
where $j_\ell$'s are the spherical Bessel functions of $\ell$-th order, $\,_{s}Y_{\ell m}(\hat{\bf k};\bf{E})$ are the spin-weighted spherical harmonics (see Appendix \ref{app:swsh} for details) and $\bf{k}$ is the momentum in Fourier space of the given wave. Moreover, in the second equalities of Eqs.~\eqref{eq:initial_conditions} we have defined
\begin{align}
\mathcal{K}^\mathbf{E}_{\ell}(k r_*)\equiv &\left( -i \right)^\ell
\left[\frac{\ell(\ell -1)}{2} \frac{j_{\ell }(kr_*)}{(kr_*)^2}
-j_{\ell }(kr_*)+\frac{j_{\ell -1}(kr_*)}{kr_*}\right]\,,
\nonumber\\
\mathcal{K}^\mathbf{B}_{\ell}(k r_*)\equiv &
\left( -i \right)^\ell
\left[(\ell +2)\frac{ j_{\ell}(kr_*)}{kr_*}-j_{\ell +1}(kr_*)\right]\,,
\end{align}
and introduced the basis for the longitudinal and circular polarizations $h_+$ and $h_\times$ given by $h_{\pm 1}\equiv \frac{1}{2}\left( h_+\mp i h_\times \right)$.

Eqs.~\eqref{eq:initial_conditions} then connect the $E/B$ modes of the tensor perturbations to the primordial gravitational waves evolved until the re-entering epoch and the latter provides the initial conditions for the evolution of the observed $E/B$ modes of the tensor perturbations until to the late-time Universe.

\section{Summary and Conclusions}
\label{sec:SC}
In this work, we have developed a formalism  to classify a set of perturbations on a Lorentzian manifold according to
their helicity rather than to their transformation properties under $SO(3)$. We have then applied this new formalism to a cosmological perturbation theory well-suited
to the structure of the light-cone previously proposed in \cite{Fanizza:2020xtv} (see also \cite{Mitsou:2020czr}).

We have first considered the entire set of linear perturbations built on top of the so-called Geodesic Light-Cone (GLC) background metric and used the gauge transformation properties for this set of variables to find what is the expression for relevant gauge invariant cosmological variables, such as the Mukhanov-Sasaki variable and the $E$- and $B$- modes of the linear tensor perturbations. Some previous attempts in this regards have been already proposed by other authors (see \cite{Frob:2021ore}). As we have proven in the Appendix~\ref{app:FL}, our results agree with the ones obtained in \cite{Frob:2021ore}. However, thanks to what we called the Scalar-PseudoScalar (SPS) decomposition, we have been able to move further and decouple the {\it electric} degrees of freedom from the {\it magnetic} ones.

Our results have been achieved thanks to the SPS decomposition and to the formalism of the spin-raising and spin-lowering operators. Indeed, the advantages in using this become manifest in order to decouple the tensor degrees of freedom from the vector and scalar ones, when mixed on the light-cone structure. This was a not trivial milestone to be achieved for two reasons. First of all, it provides a clear answer to the claim done in \cite{Mitsou:2020czr} about the physical spin-2 radiative degrees of freedom evolving along the light-cone. In fact, Eqs.~\eqref{eq:hE} and \eqref{eq:hB} precisely give what is the combination of the SPS perturbations leading to the electric and magnetic modes of the tensor perturbations, rather than just referring to their parity transformations. Secondly, this SPS decomposition of cosmological perturbations can be readily extended to the study of higher-order perturbations and then compared to other recent coordinate independent formulation of non-linear cosmological observables \cite{Ginat:2021nww}.

These two features are of particular importance when one is interested to evolve the perturbations directly along the observed light-cone rather than through constant-time hypersurfaces. 
In this framework, beyond the few analytical attempts to study the light-cone dynamics presented in \cite{Mitsou:2020czr,Buchert:2022zaa}, this work furnishes a scheme to evolve the entangled structure of the light-cone dynamics via numerical attempts, as already investigated in \cite{Tian:2021qgg}. With this in mind, having a neat, systematic classification for the relevant light-cone degrees of freedom is crucial to evolve the right variables and the results presented in this work are in line with this program. For this program to be successful, we have also discussed how to connect the light-cone variables to the primordial spectra for the case of tensor perturbations.

Given these aspects, the new formalism that we have presented
then provides the first step in the scientific program of evolving the perturbations
along the past light-cone, aiming then to address the problem of the light-cone
backreaction directly in the very same framework where light-like cosmological
observables admit simple expressions.
Whether or not this program can be achieved analytically is not clear but the recent scheme of evolving light-cone variables
numerically (see \cite{Tian:2021qgg}) seems promising in this respect. In this sense, a well-posed perturbative
framework for scalar and tensor perturbations, an explicit link with the initial conditions
as we have provided in Sect.~\ref{sec:initial} and the preliminary study of the Einstein equations
already provided in our reference \cite{Mitsou:2020czr} have put the ground basis for this program to be
pursued.

Finally, we remark that the capability to define our perturbation theory directly on the past light-cone is also relevant to connect the evolved non-linear inhomogeneities we see today to the primordial fluctuations when the perturbations are still linear. Hence, combined with the capability to evaluate the exact light-cone averages in a covariant way within the GLC gauge \cite{Fanizza:2019pfp}, the results obtained provide a step towards the evaluation of backreaction since the primordial until the late-time Universe.


\section*{Acknowledgements}
GF is thankful to the Pisa section of the INFN for the hospitality provided during the development and finalisation of this project.
GF acknowledges support by the FCT under the program {\it ``Stimulus"} with the grant no. CEECIND/04399/2017/CP1387/CT0026 and through the research project with ref. number PTDC/FIS-AST/0054/2021. GF is also member of the Gruppo Nazionale per la Fisica Matematica (GNFM) of the Istituto Nazionale di Alta Matematica (INdAM). GM and MM are supported in part by INFN under the program TAsP ({\it Theoretical Astroparticle Physics}).

\appendix

\section{Gauge transformations}
\label{app:GT}

In this appendix, we evaluate the gauge transformations of the SVT perturbations once expressed in terms of the
SPS decomposition. We will then show that these transformations agree with what is expected in the standard
perturbation theory and will then provide a sanity check of our evaluations.

We start by recalling the gauge transformations for the light-cone perturbations in Eq.~\eqref{eq:generalGLCpert}. Following
\cite{Fanizza:2020xtv}, we define the gauge field for the light-cone perturbation with $\xi^\mu\equiv\left( \xi^0,\xi^w,\hat\xi^a \right)$ and get
\begin{align}
\tilde{L} & =L+\frac{2}{a}\partial_{\tau}\xi^{w}\,,\nonumber\\
\tilde{M} & =M+\partial_{\tau}\left(\frac{\xi^0}{a}-\xi^{w}\right)+\frac{1}{a}\partial_{w}\xi^{w}+2\frac{H}{a}\xi^0\,,\nonumber\\
\tilde{N} & =N-2H\xi^0+2\partial_{w}\left(\frac{\xi^0}{a}-\xi^{w}\right)\,,\nonumber\\
\tilde{V}_{a} & =V_{a}+\frac{1}{a}\partial_{a}\xi^{w}-\bar{\gamma}_{ab}\partial_{\tau}\hat\xi^b\,,\nonumber\\
\tilde{U}_{a} & =U_{a}+\partial_{a}\left(\frac{\xi^0}{a}-\xi^{w}\right)-\bar{\gamma}_{ab}\partial_{w}\hat\xi^b\,,\nonumber\\
\tilde{\delta\gamma}_{ab} & =\delta\gamma_{ab}-2\bar{\gamma}_{ab}H\xi^0+\frac{2\bar{\gamma}_{ab}}{r}\left(\frac{\xi^0}{a}-\xi^{w}\right)-\left(\bar{\gamma}_{ac}D_{b}+\bar{\gamma}_{bc}D_{a}\right)\hat\xi^c\,.
\label{eq:GLCgaugetransf_allEqcs}
\end{align}
The latter can be further manipulated in order to give the gauge transformations for the SPS degrees of freedom. Following again \cite{Fanizza:2020xtv}, we decompose also $\hat\xi^a$ in terms of scalar and pseudo-scalar modes $\chi$ and $\hat\chi$, namely
\begin{equation}
\hat\xi^a\equiv q^{ab}\left(D_{a}\chi+\tilde{D}_{a}\hat{\chi}\right)\,,
\label{eq:chi-decomp}
\end{equation}
leading then to
\begin{align}
\tilde{\nu} =&\nu-\frac{1}{2}D^{2}\chi-\xi^0 \left( H-\frac{1}{ar} \right)-\frac{\xi^{w}}{r}\,,\nonumber\\
\tilde{v} =&v+\frac{1}{ar^{2}}\xi^{w}-\partial_{\tau}\chi\,,
&\tilde{\hat{v}}  =&\hat{v}-\partial_{\tau}\hat{\chi}\,,\nonumber\\
\tilde{u} =&u + \frac{1}{r^{2}}\left(\frac{\xi^0}{a}-\xi^{w}\right)-\partial_{w}\chi\,,
&\tilde{\hat{u}}  =&\hat{u}-\partial_{w}\hat{\chi}\,,\nonumber\\
\tilde{\mu} =&\mu-\chi\,,
&\tilde{\hat\mu}  =&\hat\mu-\hat\chi\,.
\label{eq:gaugetransf-scalarPS}
\end{align}

The relation between the SVT gauge field $\epsilon^\mu$ used in Eq.~\eqref{eq:gauge-tranform-metric} and the SPS one here introduced is then given by the finite background coordinate transformation \eqref{eq:2.18}, namely
\begin{equation}
\xi^\mu=\frac{\dd x^\mu}{\dd y^\nu}\epsilon^\nu\,,
\end{equation}
leading to
\begin{equation}
    \xi^0= a\epsilon^\eta\qquad,\qquad
    \xi^w=\epsilon^\eta+\epsilon^r\qquad,\qquad
     \hat\xi^a=\epsilon^{a}\,.
\label{eq:gaugemodes-SVT-SPS}    
\end{equation}
Moreover, for later uses, we also recall that Eqs.~\eqref{eq:2.18} gives the relations between the derivatives
\begin{equation}
\dd_\eta=a\dd_\tau +\dd_w\qquad,\qquad
\dd_r=\dd_w\qquad,\qquad
\dd_a=\dd_{a^{GLC}}\,.
\label{eq:der}
\end{equation}
These relations will be useful later to verify the correct gauge transformations of the SVT perturbations when expressed as
a linear combination of SPS fields.
Other useful relations that we provide are the following ones
\begin{align}
\Delta_{3}\epsilon = &\nabla_{i}\epsilon^{i}
= \left(\partial_{r}+\frac{2}{r}\right)\epsilon^{r}+D_{a}\epsilon^{a}
= \left(\partial_{w}+\frac{2}{r}\right)\left(\xi^{w}-\frac{\xi^0}{a}\right)+D^{2}\chi\,,\nonumber\\
\Delta_{3}\epsilon^{r} =& \left(\partial_{r}^{2}+\frac{2}{r}\partial_{r}-\frac{2}{r^{2}}+\frac{1}{r^2}D^{2}\right)\epsilon^{r}-\frac{2}{r}D_{a}\epsilon^{a}\,,\nonumber\\
    \Delta_{3}\epsilon^{a} =& \left(\partial_{r}^{2}+\frac{4}{r}\partial_{r}+\frac{1}{r^{2}}+\frac{1}{r^2}D^{2}\right)\epsilon^{a}+\frac{2}{r^{3}}D^{a}\epsilon^{r}\,.    
    \label{eq:lapl-e_r}
\end{align}

Now we can use the set of gauge transformations of the general GLC perturbations given in Eqs. \eqref{eq:GLCgaugetransf_allEqcs} and \eqref{eq:gaugetransf-scalarPS} to check the consistency of the SVT perturbations in terms of GLC perturbations. Starting with $\phi$ given in Eq. \eqref{eq:Crr}, we use Eqs.~\eqref{eq:gaugemodes-SVT-SPS} and \eqref{eq:der} to get
\begin{align}
\widetilde{\phi}  =& -\frac{1}{2}\left(a^{2}\widetilde{L}+\widetilde{N}+2a\widetilde{M}\right)\nonumber\\
=& -\frac{1}{2}\left(a^{2}L+N+2aM\right)-\left(\partial_{\tau}+\frac{1}{a}\partial_{w}\right)\xi^0
= \phi - \frac{1}{a}\partial_{\eta}\left(a\epsilon^{\eta}\right)\,, \label{eq:GT-phi}
\end{align}
which is in agreement with the expected gauge transformation for $\phi$ in Eqs.~\eqref{eq:gaugetransf}.With the use of the same equations, also the gauge transformation of $B$ can be checked. We get
(\ref{eq:scal})
\begin{align}
\Delta_{3}\widetilde{B}  = & -\left(\partial_{w}+\frac{2}{r}\right)\left(\widetilde{N}+a\widetilde{M}\right) - D^{2}\left(\widetilde{u}+a\widetilde{v}\right)\,\nonumber\\
 = & -\left(\partial_{w}+\frac{2}{r}\right)\left[\left(N+aM\right)+\left(a\partial_{\tau}+\partial_{w}\right)\left(\frac{\xi^0}{a}-\xi^{w}\right)+\partial_{w}\left(\frac{\xi^0}{a}\right)\right]\,\nonumber\\
& - D^{2}\left[u+av+\frac{\xi^0}{ar^{2}}-\left(a\partial_{\tau}+\partial_{w}\right)\chi\right]\,\nonumber\\
= & -\left(\partial_{w}+\frac{2}{r}\right)\left(N+aM\right)-D^{2}\left(u+av\right)\,\nonumber\\ 
& + \left(a\partial_{\tau}+\partial_{w}\right)\left[-\left(\partial_{w}+\frac{2}{r}\right)\left(\frac{\xi^0}{a}-\xi^{w}\right)+D^{2}\chi\right]\nonumber\\
& - \left(\partial_{w}^{2}+\frac{2}{r}\partial_{w}+r^{-2}D^{2}\right)\left(\frac{\xi^0}{a}\right)
= \Delta_{3}\left(B+\partial_{\eta}\epsilon-\epsilon^{\eta}\right)\,,
\label{eq:GT-B}
\end{align}
where on the last line we use again \eqref{eq:gaugemodes-SVT-SPS} and \eqref{eq:der}. Hence, the gauge transformation of $B$ is the one expected. On the other hand, $\psi$, as written in \eqref{eq:scal}, transforms as
\begin{align}
    \widetilde{\psi} = -\frac{1}{6}\left(\widetilde{N}+4\widetilde{\nu}\right) = & -\frac{1}{6}\left(N+4\nu\right)+H\xi^0-\frac{1}{3}\left[\left(\partial_{r}+\frac{2}{r}\right)\left(\frac{\xi^0}{a}-\xi^{w}\right)-D_{a}\hat\xi^a\right]\nonumber\\
= & \psi + \mathcal{H}\epsilon^{\eta}+\frac{1}{3}\Delta_{3}\epsilon\,.
\label{eq:GT-psi}
\end{align}

We have now all the ingredients to check directly the gauge transformation for gauge dependent curvature perturbation \eqref{eq:gaugedependentcurvature}. We preliminary derive the following gauge transformations
\begin{align}
    \frac{1}{2}\left(\frac{1}{r}\partial_{w}+\frac{1}{r^{2}}-\frac{1}{2\,r^{2}}D^{2}\right)\widetilde{N}
  =&\,\frac{1}{2}\left(\frac{1}{r}\partial_{w}+\frac{1}{r^{2}}-\frac{1}{2\,r^{2}}D^{2}\right)\left[N-2H\xi^0+2\partial_{w}\left(\frac{\xi^0}{a}-\xi^{w}\right)\right]\,\nonumber\\
  =&\,\frac{1}{2}\left(\frac{1}{r}\partial_{w}+\frac{1}{r^{2}}-\frac{1}{2\,r^{2}}D^{2}\right)\left(N-2H\xi^0\right)\,\nonumber\\
  &+\,\left(\frac{1}{r}\partial^{2}_{w}+\frac{1}{r^{2}}\partial_{w}-\frac{1}{2\,r^{2}}D^{2}\partial_{w}\right)\left(\frac{\xi^0}{a}-\xi^{w}\right)\,,\nonumber\\
  -\left(\partial^{2}_{w}+\frac{3}{r}\partial_{w}+\frac{1}{r^{2}}+\frac{1}{2\,r^{2}}D^{2}\right)\widetilde{\nu}
  =&\,-\left(\partial^{2}_{w}+\frac{3}{r}\partial_{w}+\frac{1}{r^{2}}+\frac{1}{2\,r^{2}}D^{2}\right)\left[{\nu}-H\xi^0\right.\,\nonumber\\
  &+\left.\frac{1}{r}\left(\frac{\xi^0}{a}-\xi^{w}\right)-\frac{1}{2}D^{2}\chi\right]\,,\nonumber\\
  =&\,-\left(\partial^{2}_{w}+\frac{3}{r}\partial_{w}+\frac{1}{r^{2}}+\frac{1}{2\,r^{2}}D^{2}\right)\left({\nu}-H\xi^0-\frac{1}{2}D^{2}\chi\right)\,\nonumber\\
  &-\frac{1}{r}\left(\partial^{2}_{w}+\frac{1}{r}\partial_{w}+\frac{1}{2\,r^{2}}D^{2}\right)\left(\frac{\xi^0}{a}-\xi^{w}\right)\,,\nonumber\\
   \frac{1}{2}\left(\partial_{w}+\frac{3}{r}\right)D^2\widetilde{u}
 =& \frac{1}{2}\left(\partial_{w}+\frac{3}{r}\right)D^{2}u
 +\frac{1}{2\,r^{2}}D^{2}\left[\left(\partial_{w}+\frac{1}{r}\right)\left(\frac{\xi^0}{a}-\xi^{w}\right)\right]\,\nonumber\\
 &-\frac{1}{2}\left(\partial_{w}^{2}+\frac{3}{r}\partial_{w}\right)D^{2}\chi\,,\nonumber\\
 \frac{1}{4\,r^2}D^2\left(2\widetilde{\mu} +D^{2}\widetilde{\mu}\right)=&\frac{1}{4\,r^2}D^2\left(2\mu +D^{2}\mu-2\chi-D^{2}\chi\right)\,.
 \label{eq:useful}
\end{align}
Hence, using Eqs.~\eqref{eq:useful} in the gauge transformation of $\psi$ given in Eq.~\eqref{eq:gaugedependentcurvature}, we get
\begin{align}
    \widetilde{\psi}+\frac{1}{3}\Delta_{3}\widetilde{E}
 =&\frac{1}{\Delta_{3}}\left[ \frac{1}{2}\left(\frac{1}{r}\partial_{w}+\frac{1}{r^{2}}-\frac{1}{2r^{2}}D^{2}\right)\widetilde{N}+\frac{1}{4r^{2}}\left(D^{2}\right)^{2}\widetilde{\mu}+\frac{1}{2r^{2}}D^{2}\widetilde{\mu}\right.\nonumber\\
 & \left.+\frac{1}{2}\left(\partial_{w}+\frac{3}{r}\right)D^{2}\widetilde{u}-\left(\partial_{w}^{2}+\frac{3}{r}\partial_{w}+\frac{1}{r^{2}}+\frac{1}{2r^{2}}D^{2}\right)\widetilde{\nu}\right] \,\nonumber\\
  =&\frac{1}{\Delta_{3}}\left[ \frac{1}{2}\left(\frac{1}{r}\partial_{w}+\frac{1}{r^{2}}-\frac{1}{2r^{2}}D^{2}\right)N+\frac{1}{4r^{2}}\left(D^{2}\right)^{2}\mu+\frac{1}{2r^{2}}D^{2}\mu\right.\nonumber\\
 & \left.+\frac{1}{2}\left(\partial_{w}+\frac{3}{r}\right)D^{2}u-\left(\partial_{w}^{2}+\frac{3}{r}\partial_{w}+\frac{1}{r^{2}}+\frac{1}{2r^{2}}D^{2}\right)\nu\right]+\left(H\xi^0\right)\,\nonumber\\
 =& \psi+\frac{1}{3}\Delta_{3}E+\mathcal{H}\epsilon^{\eta}\,,
 \label{eq:GT-curv-1}
\end{align}
where, once again, we have used Eqs.~\eqref{eq:gaugemodes-SVT-SPS} and \eqref{eq:der}.
Eqs.~\eqref{eq:GT-curv-1} and \eqref{eq:fluidstr} show
the gauge invariance of $\zeta,\mathcal{R}$ and $Q$, defined in
Eq.~\eqref{eq:GI-quantities} and calculated in terms of GLC perturbations
in Eqs.~\eqref{eq:GI_scalars}. Indeed, to this aim it is enough to recall that 
also from (\ref{eq:GT-psi}) and (\ref{eq:GT-curv-1}) we have that
the perturbations of an homogeneous and isotropic fluid $\bar{\rho}$ or a scalar field $\bar{\phi}$,
transform accordingly as $\widetilde{\delta\phi}=\delta\phi-\xi^0\partial_{\tau}\bar{\phi}$ and $\widetilde{\delta\rho}=\delta\rho-\xi^0\partial_{\tau}\bar{\rho}$.

With the complete set of gauge transformations for $\psi$ and $E$, we can also check the
transformation of $F_{i}$. To this end, we consider Eq.~\eqref{eq:Fi-suitable} and
then use the gauge transformations \eqref{eq:GT-psi} and
\eqref{eq:GT-curv-1}. With the use of the gauge transformations
for $\nabla^{i}C_{ir}$, i.e.
\begin{align}
\nabla^{i}\widetilde{C}_{ir} = & \left(\partial_{r}+\frac{2}{r}\right)\widetilde{N}+r^{-2}D^{2}\widetilde{u}-\frac{4}{r}\widetilde{\nu}\nonumber\\
= & \left(\partial_{r}+\frac{2}{r}\right)\left[N-2H\xi^0+2\partial_{w}\left(\frac{\xi^0}{a}-\xi^{w}\right)\right]\nonumber\\
 & +r^{-2}D^{2}u+r^{-2}D^{2}\left(\frac{\xi^0}{a}-\xi^{w}\right)-r^{-2}D^{2}\partial_{w}\chi\,\nonumber\\
 &-\frac{4}{r}\nu+\frac{2}{r}\chi+\frac{4}{r}H\xi^0-\frac{4}{r^{2}}\left(\frac{\xi^0}{a}-\xi^{w}\right)\,\nonumber\\
= & \left(\partial_{r}+\frac{2}{r}\right)N+r^{-2}D^{2}u-\frac{4}{r}\nu\nonumber\\
& -2\partial_{w}\left(H\xi^0\right)+\partial_{w}\left[\left(\partial_{w}+\frac{2}{r}\right)\left(\frac{\xi^0}{a}-\xi^{w}\right)-r^{-2}D^{2}\chi\right]\,\nonumber\\
& +\left(\partial^{2}_{w}+\frac{2}{r}\partial_{w}-\frac{2}{r^{2}}+r^{-2}D^{2}\right)\left(\frac{\xi^0}{a}-\xi^{w}\right)+\frac{2}{r}D^{2}\chi\,\nonumber\\
= & \left(\partial_{r}+\frac{2}{r}\right)N+r^{-2}D^{2}u-\frac{4}{r}\nu\,\nonumber\\ 
 &-2\partial_{w}\left(H\xi^0\right)+\Delta_{3}\left(\frac{\xi^0}{a}-\xi^{w}\right)-\partial_{w}\left(\Delta_{3}\epsilon\right)\,
\label{eq:GT-NiCir2}
\end{align}
where on the last equality we used Eq.~\eqref{eq:lapl-e_r} with $D_{a}\epsilon^{a}=D^{2}\chi$, we get
\begin{align}
    \Delta_{3}\widetilde{F}_{r}= & \nabla^{i}\widetilde{C}_{ir}+6\nabla_{i}\widetilde{\psi}-4\nabla_{i}\left(\widetilde{\psi}+\frac{1}{3}\Delta_{3}\widetilde{E}\right)\,\nonumber\\
    = & \left(\partial_{r}+\frac{2}{r}\right)N+r^{-2}D^{2}u-\frac{4}{r}\nu\,\nonumber\\ 
 &-2\partial_{w}\left(H\xi^0\right)+\Delta_{3}\left(\frac{\xi^0}{a}-\xi^{w}\right)-\partial_{w}\left(\Delta_{3}\epsilon\right)\,\nonumber\\
 &-\partial_{w}\left(N+4\nu\right)+6\partial_{w}\left(H\xi^0\right)+2\partial_{w}\left(\Delta_{3}\epsilon\right)\,\nonumber\\
 & -4\partial_{w}\frac{1}{\Delta_{3}}\left[ \frac{1}{2}\left(\frac{1}{r}\partial_{w}+\frac{1}{r^{2}}-\frac{1}{2r^{2}}D^{2}\right)N+\frac{1}{4r^{2}}\left(D^{2}\right)^{2}\mu+\frac{1}{2r^{2}}D^{2}\mu\right.\nonumber\\
 & \left.+\frac{1}{2}\left(\partial_{w}+\frac{3}{r}\right)D^{2}u-\left(\partial_{w}^{2}+\frac{3}{r}\partial_{w}+\frac{1}{r^{2}}+\frac{1}{2r^{2}}D^{2}\right)\nu\right]-4\partial_{w}\left(H\xi^0\right)\,,
 \label{eq:GTF-r-GLC}
\end{align}
which is precisely what expected since
\begin{align}
    \Delta_{3}\left(\frac{\xi^0}{a}-\xi^{w}\right)-\partial_{w}\left[\left(\partial_{w}+\frac{2}{r}\right)\left(\frac{\xi^0}{a}-\xi^{w}\right)+r^{-2}D^{2}\chi\right]=-\Delta_{3}\left(e^{r}\right)
\end{align}
with the SVT decomposition for the gauge mode $\epsilon_{r}=\partial_{r}\epsilon+e_{r}$.

With the same procedure, the gauge transformation of $\nabla^{i}\tilde{C}_{ia}$
will provide the gauge transformation of $\Delta_{3}\tilde{F}_{a}$. The former is given by
\begin{align}
\nabla^{i}\tilde{C}_{ia}= & \left(\partial_{w}+\frac{2}{r}\right)\left[r^{2}\left(D_{a}\widetilde{u}+\tilde{D}_{a}\widetilde{\hat{u}}\right)\right]+2D_{a}\widetilde{\nu}\,\nonumber\\
 & +D_{a}\left(D^{2}\widetilde{\mu}+2\widetilde{\mu}\right)+\tilde{D}_{a}\left(D^{2}\widetilde{\hat{\mu}}+\widetilde{\hat{\mu}}\right)\,\nonumber\\
= & \left(\partial_{w}+\frac{2}{r}\right)\left\{ \left[r^{2}\left(D_{a}u+\tilde{D}_{a}\hat{u}\right)\right]+D_{a}\left(\frac{\xi^0}{a}-\xi^{w}\right)-r^{2}\partial_{w}\left[D_{a}\chi+\tilde{D}_{a}\hat{\chi}\right]\right\}\,\nonumber \\
 & +2D_{a}\nu-D_{a}\left(D^{2}\chi\right)-2D_{a}\left(H\xi^0\right)+\frac{2}{r}D_{a}\left(\frac{\xi^0}{a}-\xi^{w}\right)\,\nonumber\\
 & +D_{a}\left[\left(D^{2}\mu+2\mu\right)-\left(D^{2}\chi+2\chi\right)\right]+\tilde{D}_{a}\left[\left(D^{2}\hat{\mu}+2\hat{\mu}\right)-\left(D^{2}\hat{\chi}+2\hat{\chi}\right)\right]\,\nonumber\\
= & \left(\partial_{w}+\frac{2}{r}\right)\left[r^{2}\left(D_{a}u+\tilde{D}_{a}\hat{u}\right)\right]+2D_{a}\nu+D_{a}\left(D^{2}\mu+2\mu\right)\,\nonumber\\
 & +\tilde{D}_{a}\left(D^{2}\hat{\mu}+2\hat{\mu}\right)-2D_{a}\left(H\xi^0\right)-D_{a}\left(\Delta_{3}\epsilon\right)\,\nonumber\\
 & -r^{2}\left(\partial_{w}^{2}+\frac{4}{r}\partial_{w}+\frac{1}{r^{2}}+r^{-2}D^{2}\right)\left(D_{a}\chi+\tilde{D}_{a}\hat{\chi}\right)+\frac{2}{r}D_{a}\left(\frac{\xi^0}{a}-\xi^{w}\right)\,,
 \label{eq:GT-NiC_ia}
\end{align}
where we used the relations
\begin{align}
D_{a}\left(D^{2}\chi+2\chi\right) & =D^{2}\left(D_{a}\chi\right)+D_{a}\chi\,,\nonumber\\
\tilde{D}_{a}\left(D^{2}\hat{\chi}+2\chi\right) & =D^{2}\left(\tilde{D}_{a}\hat{\chi}\right)+\tilde{D}_{a}\chi\,,
\end{align}
and, thanks to Eq.~\eqref{eq:lapl-e_r}, last line of Eq.~\eqref{eq:GT-NiC_ia} becomes
\begin{equation}
r^{2}\left(\partial_{w}^{2}+\frac{4}{r}\partial_{w}+\frac{1}{r^{2}}+r^{-2}D^{2}\right)\left(D_{a}\chi+\tilde{D}_{a}\hat{\chi}\right)+\frac{2}{r}D_{a}\left(\frac{\xi^0}{a}-\xi^{w}\right)
=\Delta_{3}\left(a^{-2}\bar{\gamma}_{ab}\hat\xi^b\right)\,.
\end{equation}
We then have that $F_a$ transforms as
\begin{align}
 \Delta_{3}\tilde{F}_{a}= & \nabla^{i}\widetilde{C}_{ia}+6\nabla_{a}\widetilde{\psi}-4\nabla_{a}\left(\widetilde{\psi}+\frac{1}{3}\Delta_{3}\widetilde{E}\right)\,\nonumber\\
= & \nabla^{i}\widetilde{C}_{ia}+6\nabla_{a}\left(N+4\nu\right)-4\nabla_{a}\left(\widetilde{\psi}+\frac{1}{3}\Delta_{3}\widetilde{E}\right)\,\nonumber\\
= & \nabla^{i}C_{ia}-2D_{a}\left(H\xi^0\right)-D_{a}\left(\Delta_{3}\epsilon\right)-\Delta_{3}\left(a^{-2}\bar{\gamma}_{ab}\hat\xi^b\right)\,\nonumber\\
& +6\nabla_{a}\left(N+4\nu\right)+6D_{a}\left(H\xi^0\right)+2D_{a}\left(\Delta_{3}\epsilon\right)\,\nonumber\\
& -4\frac{\nabla_{a}}{\Delta_{3}}\left[ \frac{1}{2}\left(\frac{1}{r}\partial_{w}+\frac{1}{r^{2}}-\frac{1}{2r^{2}}D^{2}\right)N+\frac{1}{4r^{2}}\left(D^{2}\right)^{2}\mu+\frac{1}{2r^{2}}D^{2}\mu\right.\nonumber\\
 & \left.+\frac{1}{2}\left(\partial_{w}+\frac{3}{r}\right)D^{2}u-\left(\partial_{w}^{2}+\frac{3}{r}\partial_{w}+\frac{1}{r^{2}}+\frac{1}{2r^{2}}D^{2}\right)\nu\right]-4D_{a}\left(H\xi^0\right)
 \nonumber\\
 =&\Delta_3 F_a +D_{a}\left(\Delta_{3}\epsilon\right)-\Delta_{3}\left(a^{-2}\bar{\gamma}_{ab}\hat\xi^b\right)\,,
\label{eq:GTF-a-GLC}
\end{align}
where the gauge modes in the last line of Eq.~\eqref{eq:GTF-a-GLC} can be rewritten as
\begin{equation}
D_{a}\left(\Delta_{3}\epsilon\right)-\Delta_{3}\left(a^{-2}\bar{\gamma}_{ab}\hat\xi^b\right)=-\Delta_{3}\left(a^{-2}\epsilon_{b}-D_{a}\epsilon\right)=\Delta_{3}\left(a^{-2}e_{b}\right)\,.
\end{equation}
We remark that Eqs.~\eqref{eq:GTF-r-GLC} and \eqref{eq:GTF-a-GLC} can be written in
the following compact form
\begin{equation}
\widetilde{F}_{i}^{GLC}=F_{i}^{GLC}-a^{-2}\left[\bar{f}_{i\mu}\xi^{\mu}-\partial_{i}\xi\right]\,,\label{eq:F_iGT}
\end{equation}
where the superscript (GLC) is just to indicate the expression for
$F_{i}$ in terms of GLC perturbations.

What we have shown is that GLC perturbations correctly reproduce all the gauge transformations for the scalar and vector perturbations in the SVT decomposition. We conclude this appendix by showing also the gauge invariance of the tensor perturbations in the same framework. In fact, thanks to the expected transformations for $\psi$ in Eq.~\eqref{eq:GT-psi}, for $E$ in Eq.~\eqref{eq:GT-curv-1} and for $F_{i}$ in Eq. \eqref{eq:F_iGT} we have that
\begin{align}
\tilde{h}_{rr}= & \frac{N}{2}-H\xi^0 + \partial_{w}\left(\frac{\xi^0}{a}-\xi^{w}\right)\nonumber\\ & + \left(\psi+\frac{1}{3}\Delta_{3}E\right)+H\xi^0-\nabla_{r}\left(F_{r}+\partial_{r}E\right)-\left(\frac{\xi^0}{a}-\xi^{w}\right)=h_{rr}\,,\nonumber\\
\tilde{h}_{ra}= & \frac{U_{a}}{2}-\frac{\bar{\gamma}_{ab}}{2}\partial_{w}\hat\xi^b + \frac{1}{2}\partial_{a}\left(\frac{\xi^0}{a}\right)\nonumber\\ 
  & -\frac{1}{2}\nabla_{a}\left(F_{r}+\partial_{r}E\right) + \frac{1}{2}\partial_{a}\left(\frac{\xi^0}{a}\right)
-\frac{1}{2}\nabla_{r}F_{a}+\frac{\bar{\gamma}_{ab}}{2}\partial_{w}\hat\xi^b=h_{ra}\nonumber\,,\\
  \tilde{h}_{ab}= & \frac{\delta\gamma_{ab}}{2}-\bar{\gamma}_{ab}H\xi^0 +\frac{\bar{\gamma}_{ab}}{r} \left(\frac{\xi^0}{a}-\xi^{w}\right)-\frac{1}{a^{2}}D_{(a}\xi_{b)}\nonumber\\ 
& + \bar{\gamma}_{ab}\left(\psi+\frac{1}{3}\Delta_{3}E\right)+\bar{\gamma}_{ab}H\xi^0\nonumber\\
& -\nabla_{(a}F_{b)}+\nabla_{(a}\partial_{b)}E-\frac{\bar{\gamma}_{ab}}{r} \left(\frac{\xi^0}{a}-\xi^{w}\right)+\frac{1}{a^{2}}D_{(a}\xi_{b)}=h_{ab}\,,\label{eq:gaugeinvtensor}
\end{align}
where the gauge transformation for $N$, $U_{a}$ and $\delta\gamma_{ab}$ are given in Eqs.~\eqref{eq:GLCgaugetransf_allEqcs}, the gauge dependent curvature transforms according to Eq.~\eqref{eq:GT-curv-1} and for the vector components satisfy Eqs.~\eqref{eq:GTF-r-GLC} and \eqref{eq:GTF-a-GLC}.

\section{Comparison with Fr\"ob and Lima}
\label{app:FL}
In this appendix we compare our results with the ones obtained in \cite{Frob:2021ore}. We will discuss two aspects: we first compare our gauge invariant curvature perturbation \eqref{eq:GI_scalars}
with the one provided in \cite{Frob:2021ore} where the linear GLC gauge is fixed. Afterwards, we will show that our results for the gauge invariant tensor perturbations \eqref{eq:hrr-GLC}-\eqref{eq:hab-GLC} are in agreement with the ones we can calculate with the procedure outlined in \cite{Frob:2021ore}.

\subsection{Gauge invariant curvature perturbation}
As we have already commented in the main text, although Eq.~\eqref{eq:GI_scalars} is valid in any gauge, its value coincide with the one when GLC gauge is fixed. The latter is the one provided in Eq. (5.33) of \cite{Frob:2021ore}. Since now on, we will refer to their result as $\mathcal{R}^{FL}$. In order to properly set the comparison, we preliminary point out that the definition $\mathcal{R}^{FL}$ adopted in \cite{Frob:2021ore} coincides with $2\mathcal{R}$. We then have that
\begin{align}
\Delta_{3}\mathcal{R}^{FL}  =&\frac{2H}{\partial_{\tau}\phi}\Delta_{3}\delta\varphi+\frac{2}{ar}\left(\partial_{r}+\frac{1}{r}\right)\Upsilon^{(1)}-\frac{1}{ar^{2}}D^2\Upsilon^{(1)}-\frac{1}{a^{2}r^{2}}\left(\partial_{r}+\frac{1}{r}\right)D_{a}U_{(1)}^{a}\nonumber\\
 & +\frac{1}{2a^{2}r^{4}}\left(D_{a}D_{b}\gamma_{(1)}^{ab}-D^2\gamma_{(1)}\right)-\frac{1}{2a^{2}r^{2}}\left(\partial_{r}^{2}-\frac{1}{r}\partial_{r}+\frac{1}{r^{2}}\right)\gamma_{(1)}\,,
 \label{eq:F.C.Mukhanov-Sasaki}
\end{align}
where we have already identified the inflaton perturbations $\delta\varphi$ with their $\Phi^{(1)}$, since what they call $\tau^{(1)}$ is null within the GLC gauge fixing. Moreover, the subscript and superscript $(1)$ indicate linear order expansion in Eq.~\eqref{eq:GLCmetric}. According to SPS decomposition, then, we have the following dictionary
\begin{align}
\Upsilon^{(1)}=&\frac{a}{2}N\,,\nonumber\\
U^{(1)}_a=&-a^2r^2\left( D_a u+\tilde{D}_a\hat{u} \right)\,,\nonumber\\
\gamma_{(1)}^{ab}=&2\,a^{2}r^{2}\left(\bar{q}^{ab}\nu+D^{ab}\mu+\tilde{D}^{ab}\hat{\mu}\right)\,.
\label{eq:FL-perturbations}
\end{align}
Furthermore, still following the definition given in \cite{Frob:2021ore}, we have that
\begin{equation}
\gamma_{(1)}\equiv \bar{q}^{ab}\gamma_{ab}^{(1)}=4a^{2}r^{2}\nu\,.
\label{eq:FL-g}
\end{equation}

Now we can compare term by term the two results, by using \eqref{eq:FL-perturbations} and \eqref{eq:FL-g}. We get
\begin{equation}
\begin{alignedat}{1}\frac{2}{ar}\left(\partial_{r}+\frac{1}{r}\right)\Upsilon^{(1)}-\frac{1}{ar^{2}}D^2\Upsilon^{(1)} & =\left(\frac{1}{r}\partial_{r}+\frac{1}{r^{2}}-\frac{1}{2r^{2}}D^{2}\right)N\,,\\
\frac{1}{a^{2}r^{2}}\left(\partial_{r}+\frac{1}{r}\right)D_{a}U_{(1)}^{a} & =\left(\frac{1}{r}\partial_{r}+\frac{3}{r}\right)D^{2}u\,,\\
\frac{1}{2a^{2}r^{4}}\left(D_{a}D_{b}\gamma_{(1)}^{ab}- D^2\gamma_{(1)}\right) & =-\frac{1}{r^2}\left[D^{2}\nu+D^{2}\mu+\frac{1}{2}\left(D^{2}\right)^{2}\mu\right]\,,\\
\frac{1}{2a^{2}r^{2}}\left(\partial_{r}^{2}-\frac{1}{r}\partial_{r}+\frac{1}{r^{2}}\right)\gamma_{(1)} & =2\left(\partial_{r}^{2}+\frac{3}{r}\partial_{r}+\frac{1}{r^{2}}\right)\nu\,.
\end{alignedat}
\label{eq:terms-comparison}
\end{equation}
Hence, when expressed in terms of the SPS perturbations, the result found in \cite{Frob:2021ore} becomes
\begin{align}
\Delta_{3}\mathcal{R}^{FL} =&\frac{2H}{\partial_{\tau}\bar{\varphi}}\Delta_3\delta\varphi+\left(\frac{1}{r}\partial_{r}+\frac{1}{r^{2}}-\frac{1}{2r^{2}}D^{2}\right)N+\left(\frac{1}{r}\partial_{r}+\frac{3}{r}\right)D^{2}u\nonumber\\
 & -2\left(\partial_{r}^{2}+\frac{3}{r}\partial_{r}+\frac{1}{r^{2}}+\frac{1}{2\,r^2}D^{2}\right)\nu+\frac{1}{r^2}D^{2}\mu+\frac{1}{2\,r^2}\left(D^{2}\right)^{2}\mu\,,
\label{eq:Mukh-comparison}
\end{align}
which precisely returns $\Delta_{3}\mathcal{R}^{FL}=2\Delta_{3}\mathcal{R}$. 

\subsection{Tensor perturbations}
As a further comparison between our results and \cite{Frob:2021ore},
here we discuss how to compare our results in Eqs.~\eqref{eq:hrr-GLC}-\eqref{eq:hab-GLC} for the tensor perturbations $h_{ij}$ given in \cite{Frob:2021ore}.
To this end, since in \cite{Frob:2021ore} the tensor perturbations have not been explicitly expressed in terms
of the light-cone perturbations, we first recall their approach to evaluate $h_{ij}$.

Let us start from their relation
\begin{equation}
2\left(\Delta_{3}\right)^{2}h_{ij}=\left(\Pi_{i}^{l}\Pi_{j}^{k}-\frac{1}{2}\Pi_{ij}\Pi^{lk}\right)\C_{lk}\,,
\label{eq:tensor-relation}
\end{equation}
where $\Pi_{ij}$, defined in Eq.~$\left(5.27\right)$ of \cite{Frob:2021ore}, translates to our notation as
\begin{equation}
\Pi_{ij}=\left(\bar{\gamma}_{ij}\Delta_{3}-\nabla_{i}\nabla_{j}\right)\,.
\label{eq:FC-operator}
\end{equation}
and $\C_{lk}$ is given by Eq.~\eqref{eq:BC}.

Since, for a generic scalar function $f$, we have that $\Pi^{ij}\nabla_{i}\nabla_{j}f=0$ and $\Pi^{ij}\bar{\gamma}_{ij}f=2\Delta_{3}f$, we have that
\begin{equation}
\Pi^{ij}\C_{ij}=-4\Delta_{3}\left(\psi+\frac{1}{3}\Delta_{3}E\right)\,.
\label{eq:gaugedependent-curv-op}
\end{equation}
To proceed on the evaluation of $h_{ij}$, we then explicitly write the first term in the r.h.s. of Eq.~\eqref{eq:tensor-relation} as
\begin{equation}
\Pi_{i}^{l}\Pi_{j}^{k}\C_{lk}
 =\left(\Delta_{3}\right)^{2}\C_{ij}-2\Delta_{3}\left[\nabla_{(i}\nabla^{k}\C_{j)k}\right]-\nabla_{i}\nabla_{j}\left[\left(\Pi^{lk}-\Delta_{3}\bar{\gamma}^{lk}\right)\C_{lk}\right]\,,
\label{eq:op1}
\end{equation}
and then obtain
\begin{align}
2\left(\Delta_{3}\right)^{2}h_{ij}
=&\left(\Delta_{3}\right)^{2}\C_{ij}-2\Delta_{3}\left[\nabla_{(i}\nabla^{k}\C_{j)k}\right]\nonumber\\
&-\nabla_{i}\nabla_{j}\left[\left(\frac{1}{2}\Pi^{lk}-\Delta_{3}\bar{\gamma}^{lk}\right)\C_{lk}\right]-\frac{1}{2}\Delta_{3}\left[\bar{\gamma}_{ij}\Pi^{lk}\C_{lk}\right]\,.
\label{eq:op-tensor}
\end{align}
Hence, in terms of the SPS decomposition, we get
\begin{align}
\bar{\gamma}^{ij}\C_{ij}=&N+4\nu\,,\nonumber\\
\Pi^{ij}\C_{ij} =&-2\left(\frac{1}{r}\partial_{r}+\frac{1}{r^{2}}-\frac{1}{2r^{2}}D^{2}\right)N-2\left(\partial_{r}+\frac{3}{r}\right)D^{2}u\nonumber\\
 & -\frac{1}{r^2}\left[\left(D^{2}\right)^{2}+2 D^{2}\right]\mu+4\left(\partial_{r}^{2}+\frac{3}{r}\partial_{r}+\frac{1}{r^{2}}+\frac{1}{2}r^{-2}D^{2}\right)\nu\,.
\label{eq:op-glc}
\end{align}

Finally, thanks to Eqs.~\eqref{eq:op-tensor}, \eqref{eq:op-glc} and \eqref{eq:ninjcij}, we compute all the components for the tensor perturbations according to Eq.~\eqref{eq:tensor-relation}. For $h_{rr}$, we get
\begin{align}
\left(\Delta_{3}\right)^{2}h_{rr}  =&\left(\Delta_{3}\right)^{2}\frac{N}{2}+\frac{\Delta_{3}}{2}\partial_{w}^{2}\left(N+4\nu\right)-\Delta_{3}\left\{ \partial_{w}\left[\left(\partial_{w}+\frac{2}{r}\right)N+D^{2}u-\frac{4}{r}\nu\right]\right\} \nonumber\\
 & -\frac{1}{4}\left(\Delta_{3}-\partial_{w}^{2}\right)\left[ -2\left(\frac{1}{r}\partial_{r}+\frac{1}{r^{2}}-\frac{1}{2r^{2}}D^{2}\right)N-2\left(\partial_{r}+\frac{3}{r}\right)D^{2}u-\frac{2}{r^2}D^{2}\mu\right.\nonumber\\
 & \left.-\frac{1}{r^2}\left(D^{2}\right)^{2}\mu+4\left(\partial_{r}^{2}+\frac{3}{r}\partial_{r}+\frac{1}{r^{2}}+\frac{1}{2r^2}D^{2}\right)\nu\,\right]\,.
\label{eq:FC-comparison}
\end{align}
Analogously, $h_{ra}$ is
\begin{align}
\left(\Delta_{3}\right)^{2}h_{ra} =&\frac{1}{2}\left(\Delta_{3}\right)^{2}\left[r^{2}\left(D_{a}u+\tilde{D}_a\hat{u}\right)\right]+\frac{\Delta_{3}}{2}\left[D_{a}\left(\partial_{r}-\frac{1}{r}\right)\left(N+4\nu\right)\right]\nonumber\\
 & -\left[\frac{\Delta_{3}}{2}\left(\partial_{r}-\frac{2}{r}\right)\right]\left\{ \left(\partial_{r}+\frac{2}{r}\right)\left[r^{2}\left(D_{a}u+\tilde{D}_{a}\hat{u}\right)\right]+2D_{a}\nu\right.\nonumber\\
 & \left.+D_{a}\left(D^{2}\mu+2\mu\right)+\tilde{D}_{a}\left(D^{2}\hat{\mu}+2\hat{\mu}\right)\right\} \nonumber\\
 & -\frac{\Delta_{3}}{2}\left\{ D_{a}\left[\left(\partial_{r}+\frac{2}{r}\right)N+D^{2}u-\frac{4}{r}\nu\right]\right\} \nonumber\\
 & +\frac{1}{4}D_{a}\left(\partial_{r}-\frac{2}{r}\right)\left\{ -2\left(\frac{1}{r}\partial_{r}+\frac{1}{r^{2}}-\frac{1}{2r^{2}}D^{2}\right)N-2\left(\partial_{r}+\frac{3}{r}\right)D^{2}u-\frac{2}{r^2}D^{2}\mu\right.\nonumber\\
 & \left.-\frac{1}{r^2}\left(D^{2}\right)^{2}\mu+4\left(\partial_{r}^{2}+\frac{3}{r}\partial_{r}+\frac{1}{r^{2}}+\frac{1}{2r^2}D^{2}\right)\nu\,\right\} \,.
\label{eq:FC-comparison2}
\end{align}
Finally, we have that $h_{ab}$ can be expressed as
\begin{align}
\left(\Delta_{3}\right)^{2}h_{ab} =&\left(\Delta_{3}\right)^{2}\left[r^{2}\left(\bar{q}_{ab}\nu+D_{ab}\mu+\tilde{D}_{ab}\hat{\mu}\right)\right]+\frac{\Delta_{3}}{2}\left[\left(D_{a}D_{b}+\frac{\bar{\gamma}_{ab}}{r}\partial_{w}\right)\left(N+4\nu\right)\right]\nonumber\\
 & -\Delta_{3}\left\{ \frac{\bar{\gamma}_{ab}}{r}\left[\left(\partial_{w}+\frac{2}{r}\right)N+D^{2}u-\frac{4}{r}\nu\right]+\left(\partial_{w}+\frac{2}{r}\right)\left[r^{2}\left(D_{a}D_{b}u+\tilde{D}_{ab}\hat{u}\right)\right]\right.\nonumber\\
 & \left.+D_{a}D_{b}\left(2\nu+D^{2}\mu+2\mu\right)+\tilde{D}_{ab}\left(D^{2}\hat{\mu}+2\hat{\mu}\right)\right\} \nonumber\\
 & +\frac{1}{4}\left[\bar{\gamma}_{ab}+\left(D_{a}D_{b}+\frac{\bar{\gamma}_{ab}}{r}\partial_{w}\right)\right]\left\{ -2\left(\frac{1}{r}\partial_{r}+\frac{1}{r^{2}}-\frac{1}{2r^{2}}D^{2}\right)N-\frac{2}{r^2}D^{2}\mu\right.\nonumber\\
 & \left. -\frac{1}{r^2}\left(D^{2}\right)^{2}\mu
 -2\left(\partial_{r}+\frac{3}{r}\right)D^{2}u
 +4\left(\partial_{r}^{2}+\frac{3}{r}\partial_{r}+\frac{1}{r^{2}}+\frac{1}{2r^2}D^{2}\right)\nu\,\right\} \,.
\label{eq:FC-comparison3}
\end{align}
We then have that Eqs.~\eqref{eq:FC-comparison}-\eqref{eq:FC-comparison3} respectively agree with Eqs.~\eqref{eq:hrr-GLC}-\eqref{eq:hab-GLC}, proving then the equivalence between our approach and \cite{Frob:2021ore}.

\section{Spin raising and lowering operators}
\label{app:swsh}
In order for the paper to be as self-contained as possible, 
 we provide here some useful relations of spin-weighted spherical harmonics and the spin raising and lowering operators which are extensively used in this work. This appendix is not meant to give an exhaustive discussion of the subject. It rather provides the reader with some basic useful formulas needed to obtain our results. Anyone interested in a more detailed discussion about the topic can refer to \cite{Hu:2000ee,Bernardeau:2009bm,Schmidt:2012ne,Diss-Seibert}.

Let us start by considering a function $f(\bf n)$ with spin $s$. We define the spin raising and lowering operators respectively as
\begin{align}
\ds f({\bf n}) \equiv& -\sin^s\theta\left[ \pa_\theta+\frac{i}{\sin\theta}\pa_\phi \right]\left[\sin^{-s}\theta\,f({\bf n})\right]
=-\left(\pa_\theta+\frac{i}{\sin\theta}\pa_\phi\right)f({\bf n})
+s\cot\theta f({\bf n})
\nonumber\\
\bds f({\bf n}) \equiv& -\sin^{-s}\theta\left[ \pa_\theta-\frac{i}{\sin\theta}\pa_\phi \right]\left[\sin^s\theta\,f({\bf n})\right]
=-\left(\pa_\theta-\frac{i}{\sin\theta}\pa_\phi\right)f({\bf n})
-s\cot\theta f({\bf n})\,.
\nonumber\\
\label{eq:spin_operators}
\end{align}
Here $(\theta,\phi)$ are the usual polar angles of the direction {\bf n}.
The effect of the operator $\ds$ ($\bds$) on the function $f$ is to raise (lower) its spin, such that $\ds f$ ($\bds f$) has spin $s+1$ ($s-1$). For a spin zero function $\bds f=0$. In this way, the spin-weighted spherical harmonics $_sY_{\ell m}$ are obtained from the spin zero spherical harmonics $Y_{\ell m}$ by acting with $\ds$ and $\bds$ respectively
\begin{equation}
\,_sY_{\ell m}({\bf n};{\bf E})=
\sqrt{\frac{\left(\ell-|s|\right)!}{\left(\ell+|s|\right)!}}
\begin{cases}
\ds^s Y_{\ell m}({\bf n};{\bf E})\qquad&,\qquad s\ge 0\\
\left(-1\right)^s \bds^{|s|}Y_{\ell m}({\bf n};{\bf E})\qquad&,\qquad s<0 \,.
\end{cases}
\end{equation}
The prefactor is needed to ensure that they remain normalized, i.e.
\begin{equation}
\int d\Omega_{\bf n}
\,_sY_{\ell m}({\bf n};{\bf E})
\,_sY^*_{\ell' m'}({\bf n};{\bf E})
=\delta_{\ell\ell'}\delta_{mm'}\,.
\label{eq:D3}
\end{equation}
Our notation is such that the pair ({\bf n};{\bf E}) indicates that the angles of the direction ${\bf n}$ are taken with respect to the reference frame where ${\bf E}$ is the $z$-direction. Hence, the angular derivatives in the definition of the spin operators and the dependence on ${\bf n}$ in Eq. \eqref{eq:spin_operators} are  referring to the same basis.

With reference to the basis $\e^i_\pm$ in Eq.~\eqref{eq:335}, we define $s^a_\pm\equiv ar\,\e^a_\pm$ and then write Eqs.~\eqref{eq:spin_operators} as
\begin{align}
\ds f({\bf n}) =& \left(-\sqrt{2}\,s^a_+\pa_a + s\cot\theta \right) f({\bf n})= -\sqrt{2}\,\pa_+ f({\bf n})\,,
\nonumber\\
\bds f({\bf n}) =& \left(-\sqrt{2}\,s^a_-\pa_a - s\cot\theta \right) f({\bf n})= -\sqrt{2}\,\pa_- f({\bf n})\,,
\end{align}
where $\pa_\pm\equiv s^a_\pm\pa_a\mp s\frac{\cot\theta}{\sqrt{2}}$ and
\begin{equation}
s^a_\pm \nabla_a s^c_\pm=
\frac{\cot\theta}{\sqrt{2}}\,s^c_\pm\,.
\end{equation}
For a generic field $f(\bf{n})$ with spin $s$, the following relations hold
\begin{align}
\bds\ds f({\bf n})=&D^2 f({\bf n})+s\left( 1-s\cot^2\theta \right)f({\bf n})
+2\,i\,s\frac{\cot\theta}{\sin\theta} \partial_\phi f({\bf n})\,,
\nonumber\\
\left[\bds,\ds\right]f({\bf n})=&\,2sf({\bf n})\,,
\nonumber\\
\left\{\bds,\ds\right\}f({\bf n})=&\,2\left(D^2 f({\bf n})-s^2\cot^2\theta \,f({\bf n})
+2\,i\,s\frac{\cot\theta}{\sin\theta} \partial_\phi f({\bf n})\right)\,,\nonumber\\
\bds^2\ds^2 f({\bf n})=&\,\bds\ds\bds\ds f({\bf n})
+2\left(s+1\right)\bds\ds f({\bf n})\,,\nonumber\\
\left[ \ds,\left\{ \bds,\ds \right\} \right] f({\bf n})=&-2\left( 2s+1 \right)\ds f({\bf n})\,,\nonumber\\
\left[ \bds,\left\{ \bds,\ds \right\} \right] f({\bf n})=&2\left( 2s-1 \right)\bds f({\bf n})\,.
\label{eq:C4}
\end{align}

Since they have a particular importance for the results obtained in the main text, we report some useful applications of Eqs.~\eqref{eq:C4} for $s=0,\pm 1,\pm 2$. For a spin-0 field $A(\bf{n})$, we get
\begin{align}
s^a_+s^b_+D_{ab}A({\bf n})
=&\frac{1}{2}\ds^2 A({\bf n})
\qquad,\qquad
s^a_-s^b_-D_{ab}A({\bf n})
=\frac{1}{2}\bds^2 A({\bf n})\,,
\nonumber\\
s^a_+s^b_+\widetilde{D}_{ab}A({\bf n})
=&-\frac{i}{2}\ds^2 A({\bf n})
\qquad,\qquad
s^a_-s^b_-\widetilde{D}_{ab}A({\bf n})
=\frac{i}{2}\bds^2 A({\bf n})\,,
\nonumber\\
\bds^2\ds^2 A({\bf n})=&\,\left( D^2 \right)^2 A({\bf n})
+2D^2 A({\bf n})
\label{eq:useful2}
\end{align}
where we have used $s^b_\pm \epsilon^a_b=\mp i s^a_\pm$.
For a spin-1 vector field $S_i({\bf n})$, we have that
\begin{equation}
s^a_\pm D^2 S_a({\bf n}) =\frac{1}{2}\left\{ \bds,\ds \right\}S_\pm({\bf n})\,,
\end{equation}
and for a spin-2 tensor field $S_{ij}({\bf n})$, we also have
\begin{equation}
s^a_\pm s^b_\pm D^2 S_{ab}({\bf n}) =\frac{1}{2}\left\{ \bds,\ds \right\}S_{\pm\pm}({\bf n})\,.
\end{equation}

\section{Derivations of the {\it E} and {\it B} modes for tensor perturbations}
\label{app:EB}
In this appendix we report the derivation of the $E$- and $B$- modes of the helicity-2 radiative degrees of freedom, reported in the main text in Eqs.~\eqref{eq:hE} and \eqref{eq:hB}. To this aim, let us first consider the field $S_i$ needed to evaluate Eq.~\eqref{eq:tensor-projections}
\begin{equation}
\Delta_{3}S_{i}=\Delta_{3}\left(F_{i}+\partial_{i}E\right)=\nabla^{j}\C_{ji}+\nabla_{i}\left[2\left(\psi+\frac{1}{3}\Delta_{3}\right)-\Delta_{3}E\right]\,.\label{eq:Si-1}
\end{equation}
Hence, in order to compute the $E$- and $B$- modes of $S_i$, we consider the first term on the r.h.s. of Eq.~\eqref{eq:Si-1} given by
\begin{equation}
\nabla^{j}\C_{ji}=a^{2}\left(\n^l \n^k+2\,\e^{(l}_+\e^{k)}_-\right)\nabla_{l}\C_{ki}\,,
\label{eq:D2}
\end{equation}
where we have rewritten the background $3$-D metric as $\bar g^{ij}=\n^i \n^j+2\,\e^{(i}_+\e^{j)}_-$. In this way, the three terms on the r.h.s. of Eq.~\eqref{eq:D2} projected onto $\e^i_\pm$ are given by
\begin{align}
a^{2}\bar{e}_{\pm}^{i}\bar{n}^{l}\bar{n}^{k}\nabla_{l}\C_{ki}=&\,a\partial_{\rVert}\C_{\rVert\pm}=\frac{1}{a}\partial_{\rVert}\mathcal{T}_{\rVert\pm}\,,\nonumber \\
a^{2}\bar{e}_{\pm}^{i}\bar{e}_{\pm}^{l}\bar{e}_{\mp}^{k}\nabla_{l}\C_{ki}= &\,\frac{a}{r}\left(\partial_{\pm}\C_{\pm\mp}+\C_{\rVert\pm}\right)=\frac{1}{ar}\mathcal{T}_{\rVert\pm}+\frac{1}{ar}\partial_{\pm}\mathcal{T}\,,\nonumber \\
a^{2}\bar{e}_{\pm}^{i}\bar{e}_{\mp}^{l}\bar{e}_{\pm}^{k}\nabla_{l}C_{ki}= &\,\frac{a}{r}\left(\partial_{\mp}\C_{\pm\pm}+2\,\C_{\pm\rVert}\right)=\frac{2}{ar}\mathcal{T}_{\rVert\pm}+\frac{1}{ar}\partial_{\mp}\mathcal{T}_{\pm\pm}\,,
\label{eq:projections}
\end{align}
where we have defined $\pa_\rVert\equiv \n^i\pa_i$.

Using Eqs.~\eqref{eq:projections} and applying Eqs.~\eqref{eq:proj} and \eqref{eq:45}, we then obtain
\begin{align}
\left(\e_{\pm}^{i}\nabla^{j}\C_{ij}\right)^{\mathbf{E}}= & \frac{1}{a}\left(\partial_{r}+\frac{3}{r}\right)\mathcal{T}_{\rVert}^{\mathbf{E}}-\frac{1}{ar\sqrt{2}}\mathcal{T}^{\mathbf{E}}-\frac{D^{2}}{ar\sqrt{2}}\mathcal{T}\,,
\nonumber\\
\left(\e^i_\pm\nabla^{j}\C_{ji}\right)^{\mathbf{B}}=&\frac{1}{a}\left(\partial_{\rVert}+\frac{3}{r}\right)\mathcal{T}_{\rVert}^{\mathbf{B}}-\frac{1}{ar\sqrt{2}}\mathcal{T}^{\mathbf{B}}\,,
\label{eq:F-B}
\end{align}
where all the $\T$ quantities are given in Sect.~\ref{sec:gen}. At this point, we obtain that the $E/B$ modes of $S_i$ are given by
\begin{align}
\left(\bar{e}_{\pm}^{i}\Delta_{3}S_{i}\right)^{\mathbf{E}}
= &\,\Delta_{3}S^{\mathbf{E}}-\frac{2}{ar^{2}\sqrt{2}}D^{2}\left(\partial_{r}E+F_{r}\right)
\nonumber\\
= & -\frac{1}{a\sqrt{2}}\left(\partial_{w}+\frac{3}{r}\right)\left(rD^{2}u\right)
-\frac{1}{ar\sqrt{2}}\left(D^{2}+2\right)D^{2}\mu
-\frac{2D^{2}}{ar\sqrt{2}}\nu
\nonumber \\
 & -\frac{D^{2}}{ar\sqrt{2}}\left[2\left(\psi+\frac{1}{3}\Delta_{3}E\right)-\Delta_{3}E\right]\,,
\nonumber\\
\left(\bar{e}_{\pm}^{i}\Delta_{3}S_{i}\right)^{\mathbf{B}}= & -\frac{1}{a\sqrt{2}}\left(\partial_{w}+\frac{3}{r}\right)\left(rD^{2}\hat{u}\right)-\frac{1}{ar\sqrt{2}}\left(D^{2}+2\right)D^{2}\hat{\mu}
\end{align}
where we have used $\left(\bar{e}_{\pm}^{i}\Delta_{3}S_{i}\right)^{\mathbf{E}}=\Delta_{3}S^{\mathbf{E}}-\frac{2}{r^{2}\sqrt{2}}D^{2}S_\rVert$, with $S_\rVert\equiv \n^i S_i$, while $E$, $F_r$ and the gauge dependent curvature perturbation $\psi+\frac{1}{3}\Delta_{3}E$ are given in Eqs.~\eqref{eq:scal}, \eqref{eq:Fa} and \eqref{eq:gaugedependentcurvature}. Finally, the final expressions of the $E/B$ modes of the radiative degrees of freedom in terms of the SPS perturbations are given by
\begin{align}
h^{\mathbf{E}}= & \frac{\left(D^{2}+2\right)}{2a^{2}}\left\{ D^{2}\mu-\frac{1}{r\Delta_{3}}\left[\left(\partial_{w}+\frac{3}{r}\right)\left(rD^{2}u\right)+\frac{\left(D^{2}+2\right)}{r}D^{2}\mu+\frac{2D^{2}}{r}\nu\right]\right\} \nonumber \\
 & +\frac{\left(D^{2}+2\right)}{2a^{2}r}\frac{1}{\Delta_{3}}\left\{ \frac{D^{2}}{r}\left\{ \frac{1}{3}\left(-\frac{2}{r}\partial_{w}+\Delta_{3}\right)\left\{ \frac{1}{\Delta^2_{3}}\left[\frac{1}{2}\left(\partial_{w}^{2}+\frac{5}{r}\partial_{w}+\frac{3}{r^{2}}\right)\left(N-2\nu\right)\right.\right.\right.\right.\nonumber \\
 & \left.\left.-\frac{1}{4\,r^{2}}D^{2}\left(N+4\nu\right)+\frac{1}{6\,r^{2}}\left(\partial_{w}+\frac{1}{r}\right)D^{2}u+\frac{3}{2\,r^{2}}D^{2}\nu
 +\frac{3}{4r^2}\left(D^2+2\right)D^2\mu\right]\right\} \nonumber \\
 & \left.\left.+\frac{2}{r}\frac{1}{\Delta_{3}}\left[\left(\frac{1}{3}\partial_{w}+\frac{1}{r}\right)\left(2N-4\nu\right)+D^{2}u\right]-\frac{1}{3}\left(N+4\nu\right)\right\} \right\}\,,
\end{align}
and
\begin{align}
h^{\mathbf{B}}= & \frac{\left(D^{2}+2\right)}{2a^2}\left\{ D^{2}\hat{\mu}-\frac{1}{r\Delta_{3}}\left[\left(\partial_w+\frac{3}{r}\right)\left(rD^{2}\hat{u}\right)+\frac{\left(D^{2}+2\right)}{r}D^{2}\hat{\mu}\right]\right\}\,.
\end{align}

\bibliographystyle{JHEP}
\bibliography{Ref-GLCPert-SVT-GLC}

\end{document}